\documentclass[a4paper]{cas-sc}

\usepackage{apacite}
\usepackage[authoryear]{natbib}
\usepackage{amsmath}
\usepackage{mathtools}
\usepackage{enumitem}
\usepackage{bm}
\usepackage[ruled,linesnumbered]{algorithm2e}
\usepackage{algcompatible}
\SetKw{Kwin}{$\in$}
\SetKw{Kwpara}{in parallel}
\SetKw{KwOp}{Output:} 
\SetKwRepeat{Do}{do}{while}
\newlength\myindent
\makeatletter
\newcommand{\nosemic}{\renewcommand{\@endalgocfline}{\relax}}
\newcommand{\dosemic}{\renewcommand{\@endalgocfline}{\algocf@endline}}
\makeatother
\usepackage{wrapfig}
\usepackage[figuresright]{rotating}
\usepackage{amssymb}
\usepackage{varioref}
\usepackage{hyperref}
\usepackage{cleveref}

\newcommand{\pluseq}{\mathrel{+}=}
\usepackage{nomencl}
\usepackage{multirow}
\usepackage{ragged2e}
\usepackage{graphicx}
\usepackage{textcomp}
\usepackage{lipsum}
\usepackage{xcolor}
\usepackage{array, booktabs, colortbl, tabularx}
\usepackage{longtable}
\usepackage{pdflscape}
\usepackage{afterpage}
\usepackage{stfloats}
\usepackage{algpseudocode}
\usepackage{wrapfig}
\usepackage{subcaption}
\usepackage{caption}
\usepackage{array}
\newcolumntype{P}[1]{>{\centering\arraybackslash}p{#1}}
\newcolumntype{M}[1]{>{\centering\arraybackslash}m{#1}}
\makenomenclature

\newcommand*\oline[1]{%
\vbox{%
	\hrule height 0.5pt
	\kern0.35ex
	\hbox{%
		\kern-0.1em
		\ifmmode#1\else\ensuremath{#1}\fi
		\kern-0.1em
	}
}
}

\usepackage{tikz}
\usetikzlibrary{decorations.pathreplacing,calc}

\usepackage{amsthm}
\theoremstyle{definition}

\theoremstyle{remark}

\ExplSyntaxOn
\cs_gset_protected:Npn \__make_fig_caption:nn #1#2
{
\skip_vertical:N \l_fig_abovecap_skip
\parbox { \dim_eval:n { \l_fig_width_dim } }
{
	\tl_use:N \l_fig_align_tl
	\sffamily \small \textbf{\color{scolor}#1:}~#2\par
}
\skip_vertical:N \l_fig_belowcap_skip
}
\cs_gset_protected:Npn \__make_tbl_caption:nn #1#2
{
\skip_vertical:N \l_tbl_abovecap_skip
\parbox{ \dim_eval:n { \l_tbl_width_dim } }
{
	\tl_use:N \l_tbl_align_tl
	\sffamily \small \textbf{\color{scolor}#1}\par#2\par\vskip4pt
}
\skip_vertical:N \l_tbl_belowcap_skip
}
\ExplSyntaxOff

\def\tsc#1{\csdef{#1}{\textsc{\lowercase{#1}}\xspace}}
\tsc{WGM}
\tsc{QE}
\tsc{EP}
\tsc{PMS}
\tsc{BEC}
\tsc{DE}

\begin{document}
\shorttitle{Crew pairing optimization framework for tackling large-scale \& complex flight networks}
\shortauthors{D. Aggarwal et~al.} 

\title [mode = title]{Airline Crew Pairing Optimization Framework for Large Networks with Multiple Crew Bases and Hub-and-Spoke Subnetworks} 

\author[1]{Divyam Aggarwal}[orcid=0000-0003-0740-780X]
\ead{daggarwal@me.iitr.ac.in}

\address[1]{Department of Mechanical \& Industrial Engineering (MIED), Indian Institute of Technology Roorkee, Roorkee, Uttarakhand-247667, India}

\author[1]{Dhish Kumar Saxena}[orcid=0000-0001-7809-7744]
\cormark[1]
\ead{dhish.saxena@me.iitr.ac.in}

\author[2]{Thomas B\"ack}[orcid=0000-0001-6768-1478]
\ead{t.h.w.baeck@liacs.leidenuniv.nl}

\address[2]{Leiden Institute of Advanced Computer Science (LIACS), Leiden University, Niels Bohrweg 1, 2333 CA Leiden, the Netherlands}

\author[2]{Michael Emmerich}[orcid=0000-0002-7342-2090]
\ead{m.t.m.emmerich@liacs.leidenuniv.nl}

\cortext[cor1]{\textit{Corresponding author; Email Address: dhish.saxena@me.iitr.ac.in; Postal Address: Room No.-231, East Block, MIED, IIT Roorkee, Roorkee, Uttarakhand-247667, India; Phone: +91-8218612326}}

\begin{abstract}
	\textit{Crew Pairing Optimization} aims at generating a set of flight sequences (\textit{crew pairings}), covering \textit{all} flights in an airlines' flight schedule, at \textit{minimum cost}, while satisfying several \textit{legality} constraints. CPO is critically important for airlines' business viability considering that the crew operating cost is second only to the fuel cost. It poses an NP-hard combinatorial optimization problem, to tackle which, the state-of-the-art relies on relaxing the underlying Integer Programming Problem (IPP) into a Linear Programming Problem (LPP),  solving the latter through Column Generation (CG) technique, and  integerization of the resulting LPP solution. However, with the growing scale and complexity of the airlines' networks (those with large number of flights, multiple crew bases and/or multiple hub-and-spoke subnetworks), the efficacy of the conventionally used \textit{exact} CG-implementations is severely marred, and their utility has become questionable. This paper proposes an Airline Crew Pairing Optimization Framework, $AirCROP$, whose \textit{constitutive modules} include the Legal Crew Pairing Generator, Initial Feasible Solution Generator, and an Optimization Engine built on heuristic-based CG-implementation. $AirCROP$'s novelty lies in not just \textit{the design of its constitutive modules} but also in \textit{how these modules interact}. In that, insights in to several important questions which the literature is otherwise silent on, have been shared. These relate to sensitivity analysis of $AirCROP's$ performance in terms of final solutions' cost quality and run-time, with respect to - sources of variability over multiple runs for a given problem; cost quality of the initial solution and the run-time spent to obtain it;
and termination parameters for LPP-solutioning and IPP-solutioning. In addition, the efficacy of the $AirCROP$ has been: (a) demonstrated on real-world airline flight networks with an unprecedented conjunct scale-and-complexity, marked by over 4200 flights, 15 crew bases, and billion-plus pairings, and (b) validated by the research consortium’s industrial sponsor. It is hoped that with the emergent trend of conjunct scale and complexity of airline networks, this paper shall serve as an important milestone for affiliated research and applications.
\end{abstract}

\begin{keywords}
	Airline Crew Scheduling \sep Crew Pairing \sep Combinatorial Optimization \sep Column Generation \sep Mathematical Programming \sep Heuristics
\end{keywords}

\maketitle

\section{Introduction} \label{sec:intro}
\par Airline scheduling poses some of the most challenging optimization problems encountered in the entire Operations Research (OR) domain. For a large-scale airline, the crew operating cost constitutes the second-largest cost component, next to the fuel cost, and even its marginal improvements may translate to annual savings worth millions of dollars. Given the potential for huge cost-savings, Airline Crew Scheduling is recognized as a critical planning activity. It has received an unprecedented attention from the researchers of the OR community over the last three decades. Conventionally, it is tackled by solving two problems, namely, \textit{Crew Pairing Optimization Problem} (CPOP) and \textit{Crew Assignment Problem}, in a sequential manner. The former problem is aimed at generating a set of flight sequences (each called a \textit{crew pairing}) that covers all flights from an airlines' flight schedule, at minimum cost, while satisfying several legality constraints linked to federations' rules, labor laws, airline-specific regulations, etc. These optimally-derived crew pairings are then fed as input to the latter problem, which is aimed to generate a set of pairing sequences (each sequence is a \textit{schedule} for an individual crew member), while satisfying the corresponding crew requirements. Being the foremost step of the airline crew scheduling, CPOP is the main focus of this paper, and interested readers are referred to \cite{barnhart2003airline} for a comprehensive review of the airline crew scheduling.
\par CPOP is an \textit{NP-hard}\footnote{For NP-hard (NP-complete) problems, no polynomial time algorithms on sequential computers are known up to now. However, verification of a solution might be (can be) accomplished efficiently, i.e., in polynomial time.} combinatorial optimization problem \citep{garey2002computers}. It is modeled as either a \textit{set partitioning problem} (SPP) in which each flight is allowed to be covered by only one pairing, or a \textit{set covering problem} (SCP) in which each flight is allowed to be covered by more than one pairing. In CPOP, a crew pairing has to satisfy hundreds of legality constraints (Section~\ref{sec:constraints}) to be classified as \textit{legal}, and it is imperative to generate legal pairings in a time-efficient manner to assist optimization search. Several legal pairing generation approaches, based on either a flight- or a duty-network, have been proposed in the literature \citep{aggarwal2018large}. Depending upon how the legal pairing generation module is invoked, two CPOP solution-architectures are possible. In the first architecture, all possible legal pairings are enumerated \textit{a priori} the CPOP-solutioning. However, this is computationally-tractable only for small-scale CPOPs (with $\approx$<1000 flights). Alternatively, legal pairings are generated during each iteration of the CPOP-solutioning, but only for a subset of flights, so the CPOP solution could be partially improved before triggering the next iteration. Such an architecture mostly suits medium- to large-scale CPOPs (with $\approx \geq$1000 flights) involving millions/billions of legal pairings, whose complete-enumeration is computationally-intractable. 

In terms of solution-methodologies, \textit{heuristic-based} optimization techniques and \textit{mathematical programming} techniques, are commonly employed (Section~\ref{sec:relatedwork}). In the former category, \textit{Genetic Algorithms} (GAs) which are population-based randomized-search heuristics \citep{goldberg2006genetic} are most commonly used. However, they are found to be efficient only for tackling very small-scale CPOPs \citep{ozdemir2001flight}. Alternatively, several mathematical programming based approaches do exist to solve CPOPs of varying-scales. CPOP is inherently an Integer Programming Problem (IPP), and some approaches have used standard Integer Programming (IP) techniques to find a best-cost pairing subset from a pre-enumerated pairings' set \citep{hoffman1993solving}. However, these approaches have proven effective only with small-scale CPOPs with up to a million pairings. This perhaps explains the prevalence of an altogether different strategy, in which the original CPOP/IPP is relaxed into a Linear Programming Problem (LPP); the LPP is solved iteratively by invoking a LP solver and relying on \textit{Column Generation} (CG) technique to generate new pairings as part of the pricing sub-problem; and finally, the resulting LPP solution is integerized using IP techniques and/or some special \textit{connection-fixing} heuristics. The challenge associated with this strategy is that even though the LPP solver may lead to a near-optimal LPP solution, the scope of finding a good-cost IPP solution is limited to the pairings available in the LPP solution. To counter this challenge, heuristic implementations of \textit{branch-and-price} framework (\cite{barnhart1998branch}) in which CG is utilized during the integerization phase too, have been employed to generate new legal pairings at nodes of the IP-search tree. However, the efficacy of such heuristic implementations depends on a significant number of algorithmic-design choices (say, which branching scheme to adopt, or how many CG-iterations to perform at the nodes). Furthermore, it is noteworthy that the scale and complexity of flight networks have grown alarmingly over the past decades. As a result, an inestimably large number of new pairings are possible under the pricing sub-problem, given which most existing solution methodologies are rendered  computationally-inefficient. Recognition of such challenges have paved the way towards domain-knowledge driven CG strategies to generate a manageable, yet crucial part of the overall pairings’ space under the pricing sub-problem \citep{zeren2016novel}. Though rich in promise, the efficacy of this approach is yet to be explored vis-$\grave{a}$-vis the emergent large-scale and \textit{complex} flight networks characterized by \textit{multiple crew bases} and/or \textit{multiple hub-and-spoke subnetworks} where billions of legal pairings are possible. 

\par In an endeavor to address airline networks with conjunct scale and complexity, this paper proposes an Airline Crew Pairing Optimization Framework ($AirCROP$) based on domain-knowledge driven CG strategies, and:
\begin{itemize}
	\item presents not just \textit{the design of its constitutive modules} (including Legal Crew Pairing Generator, Initial Feasible Solution Generator, and Optimization Engine powered by CG-driven LPP-solutioning and IPP-solutioning), but also \textit{how these modules interact}
	\item discusses \textit{how sensitive its performance is} to - sources of variability over multiple runs for a given problem; cost quality of the initial solution and the run-time spent to obtain it; 
and termination parameters for LPP-solutioning and IPP-solutioning. Such an investigation promises important insights for researchers and practitioners on critical issues which are otherwise not discussed in the existing literature.
	\item presents empirical results for real-world, large-scale (over 4200 flights), complex flight network (over 15 crew bases and multiple hub-and-spoke subnetworks) for a US-based airline, the data for which has been provided by the research consortium’s industrial partner.
\end{itemize}
\par The outline of the remaining paper is as follows. Section~\ref{sec:ACP} discusses the underlying concepts, related work, and problem formulation; Section~\ref{sec:aircrop} entails the details of the proposed $AirCROP$; Section~\ref{sec:exp} presents the results of the computational experiments along with the corresponding observations; and Section~\ref{sec:conc} concludes the paper as well as briefly describes the potential future directions.

\section{Crew Pairing Optimization: Preliminaries, Related Work and Problem Formulation} \label{sec:ACP}
This section first describes the preliminaries, including the associated terminology, pairings' legality constraints, and pairings' costing criterion. Subsequently, the related work is presented in which the existing CPOP solution approaches are discussed. Lastly, the airline CPOP formulation is presented.

\subsection{Associated Terminology}
\begin{figure}[pos=htbp,align=\centering]
	\includegraphics[width=0.65\columnwidth, keepaspectratio]{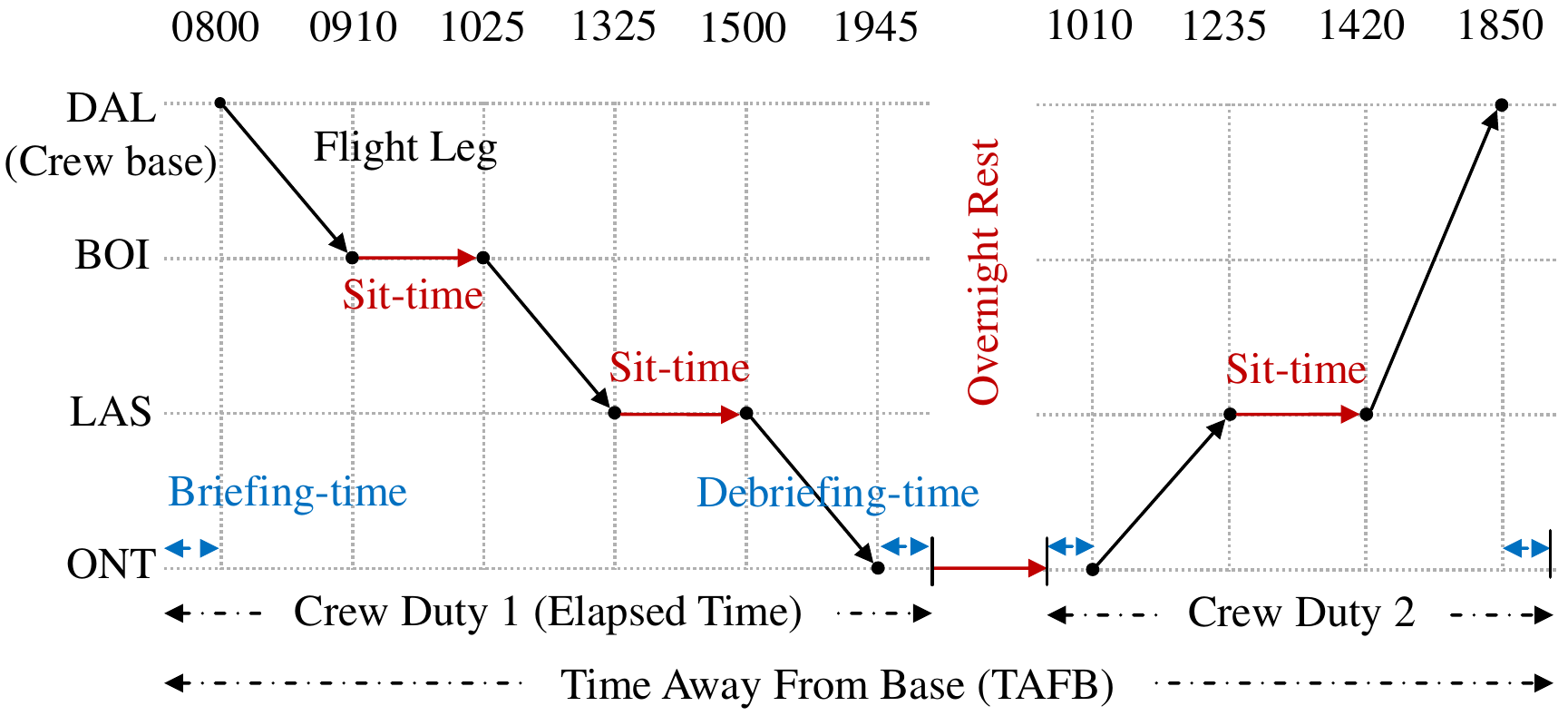}
	\caption{An example of a crew pairing starting from \textit{Dallas} (DAL) crew base}
	\label{fig:pairing}
\end{figure}

In airline crew operations, each crew member is assigned a fixed (home) airport, called a \textit{crew base}. A \textit{crew pairing} (or a \textit{pairing}) is a flight sequence operated by a crew, that begins and ends at the same crew base, and satisfies the given pairing legality constraints (detailed in Section~\ref{sec:constraints}). An example of a crew pairing with the Dallas (DAL) airport as the crew base is illustrated in Figure~\ref{fig:pairing}. In a crew pairing, the legal sequence of flights operated by a crew in a single working day (not necessarily equivalent to a calendar day) is called a \textit{crew duty} or a \textit{duty}. A \textit{sit-time} or a \textit{connection-time} is a small rest-period, provided between any two consecutive flights within a duty for facilitating operational requirements such as aircraft changes by the crew, turn-around operation for the aircraft, etc. An \textit{overnight-rest} is a longer rest-period, provided between any two consecutive duties within a pairing. Moreover, two short-periods, provided in the beginning and ending of any duty within a pairing, are called \textit{briefing} and \textit{de-briefing time}, respectively. The total time elapsed in a crew pairing, i.e., the time for which a crew is away from its crew base is called the \textit{time away from base} (TAFB).
Sometimes, it is required for a crew to be transported at an airport to fly their next flight. For this, the crew travels as passenger in another flight, flown by another crew, to arrive at the required airport. Such a flight is called a \textit{deadhead flight} or a \textit{deadhead} for the crew traveling as passenger. It is desired by an airline to minimize the number of deadheads (ideally zero), as it affects the airline's profit in two-folds. Firstly, the airline suffers a loss of the revenue on the passenger seat being occupied by the deadhead-ing crew, and secondly, the airline has to pay the hourly wages to the deadhead-ing crew even when it is not operating the flight.

\subsection{Crew Pairing: Legality Constraints and Costing Criterion} \label{sec:constraints}
To govern the safety of crew members, airline federations such as European Aviation Safety Agency, Federal Aviation Administration, and others, have laid down several rules and regulations, which in addition to the airline-specific regulations, labor laws, etc. are required to be satisfied by a pairing to be ``legal''. These legality constraints could be broadly categorized as follows:
\begin{itemize}
	\item \textit{Connection-city constraint} ($\mathcal{C}_{connect}$): this constraint requires the arrival airport of a flight (or the last flight of a duty) within a pairing to be same as the departure airport of its next flight (or the first flight of its next duty).
	\item \textit{Sit-time ($C_{sit}$) and Overnight-rest ($\mathcal{C}_{night}$) constraints}: these constraints imposes the respective maximum and minimum limits on the duration of sit-times and overnight-rests, where these limits are governed by airlines and federations' regulations.
	\item \textit{Duty constraints} ($\mathcal{C}_{duty}$): these constraints govern the regulations linked to the crew duties. For instance, they impose maximum limits on the-- number of flights allowed in a duty of a pairing; duty elapsed-time and the corresponding flying-time; number of duties allowed in a pairing, etc.
	\item \textit{Start- and end-city constraint} ($\mathcal{C}_{base}$): this constraint requires the beginning airport (departure airport of the first flight) and ending airport (arrival airport of the last flight) of a pairing, to be the same crew base.
	\item \textit{Other constraints} ($\mathcal{C}_{other}$): Airlines formulate some specific constraints, according to their operational requirements, so as to maximize their crew utilization. For example, a pairing is refrained from involving overnight-rests at the airports that belong to the same city as the crew base from which the pairing started, etc.
\end{itemize}
Considering the multiplicity of the above constraints, it is critical to develop a time-efficient \textit{legal crew pairing generation approach}, enabling their prompt availability, when their requirement arises during the optimization.
\par In general, a pairing's cost could be split into the \textit{flying cost} and \textit{non-flying (variable) cost}. The flying cost is the cost incurred in actually flying all the given flights, and is computed on hourly-basis. The variable cost is the cost incurred during the non-flying hours of the pairing, and is made up of two sub-components, namely, \textit{hard cost} and \textit{soft cost}. The hard cost involves the pairing's hotel cost, meal cost, and  \textit{excess pay}-- the cost associated with the difference between the guaranteed hours of pay and the actual flying hours. Here, the pairing's hotel cost is the lodging cost incurred during its overnight-rests, and its meal cost is computed as a fraction of its TAFB. The soft cost is the undesirable cost associated with the number of aircraft changes (during flight-connections) in the pairing, etc.

\subsection{Related Work} \label{sec:relatedwork}
As mentioned in Section~\ref{sec:intro}, the existing CPOP solution approaches are based on either heuristic or mathematical programming techniques. Among the heuristic-based approaches, GA is the most widely adopted technique, and \cite{beasley1996genetic} is the first instance to customize a GA (using guided GA-operators) for solving a general class of SCPs. In that, the authors validated their proposed approach on small-scale synthetic test cases (with over 1,000 rows and just 10,000 columns). The important details of the GA-based CPOP solution approaches, available in the literature, are reported in Table~\ref{tab:GAlitRev}.
\begin{table}[pos=htbp,align=\centering]
	\small
	\centering \caption{Key facts around the GA-based CPOP solution approaches, available in the literature}
	\begin{center}
		\begin{tabular}{ccccc}
			\toprule
			\textbf{Literature Studies} & \textbf{Modeling} & \textbf{Timetable} & \textbf{Airline Test Cases*} & \textbf{Airlines}\\
			\midrule
			\cite{levine1996application} & Set Partitioning & - & 40R; 823F; 43,749P & - \\
			\cite{ozdemir2001flight} & Set Covering & Daily & 28R; 380F; 21,308P & Multiple Airlines \\
			\cite{kornilakis2002crew} & Set Covering & Monthly & 1R; 2,100F; 11,981P & Olympic Airways \\
			\cite{zeren2012improved} & Set Covering & Monthly & 1R; 710F; 3,308P & Turkish Airlines \\
			\cite{deveci2018evolutionary} & Set Covering & - & 12R; 714F; 43,091P & Turkish Airlines \\
			\bottomrule
		\end{tabular}
		\\
		R represents the number of real-world test cases considered; F and P represents the maximum number of flights and pairings covered, therein. 
	\end{center}
	\label{tab:GAlitRev}
\end{table}
Notably, the utility of the studies reported in the table, have been demonstrated on CPOPs with reasonably small number of flights, leading to relatively smaller number of pairings. Though, CPOPs with 2,100 and 710 flights have been tackled by \citet{kornilakis2002crew} and \citet{zeren2012improved} respectively, only a subset of all possible legal pairings has been considered by them for finding the reported solutions. \citet{zeren2012improved} proposed a GA with highly-customized operators, which efficiently solved small-scale CPOPs but failed to solve large-scale CPOPs with the same search-efficiency. Furthermore, \cite{aggarwal2020realworld} tackled a small-scale CPOP (with 839 flights and multiple hub-and-spoke sub-networks) using a customized-GA (with guided operators) as well as mathematical programming techniques. The authors concluded that customized-GAs are inefficient in solving complex versions of even small-scale flight networks, compared to a mathematical programming-based solution approach.
\par Several mathematical programming-based CPOP solution approaches have been proposed in the literature over past few decades, and based on the size and characteristics of the flight network being tackled, these approaches have been categorized into either of the three general classes. In the first class of approaches, all legal pairings or a subset of good pairings are enumerated prior to the CPOP-solutioning, and the corresponding CPOP/IPP model is solved using standard IP techniques (such as \textit{branch-and-bound} algorithm \citep{land1960an}). \cite{gershkoff1989optimizing} proposed an iterative solution approach, which is initialized using a set of artificial pairings (each covering a single flight at a high pseudo-cost). In that, each iteration involves selection of very few pairings (5 to 10); enumeration of all legal pairings using the flights covered in the selected pairings; optimization of the resulting SPP to find the optimal pairings; and lastly, replacement of the originally selected pairings with the optimal pairings, only if the latter offers a better cost. The search-efficiency of such an approach is highly dependent on the sub-problem-size (handled up to 100 flights and 5,000 pairings), as the length and breadth of the branching tree increases drastically with an increase in  sub-problem-size. \cite{hoffman1993solving} proposed an alternative approach to tackle SPPs with up to 825 flights and 1.05 million pairings in which all possible pairings are enumerated a priori, and the resulting SPP is solved to optimality using a \textit{branch-and-cut} algorithm\footnote{The branch-and-cut algorithm was first proposed by \cite{padberg1991branch} to solve Mixed Integer Programs (MIP), by integrating the standard \textit{branch-and-bound} and \textit{cutting-plane} algorithms. For comprehensive details of the MIP solvers, interested readers are referred to \cite{lodi2009mixed, linderoth2010milp, achterberg2013mixed}.}. Such approaches are efficient only in tackling small-scale CPOPs, that too with up to a million pairings. However, even small-scale CPOPs may involve large number of pairings (an instance reported in \cite{vance1997heuristic} had 250 flights and over five million pairings), rendering it computationally-intractable to use such approaches.
\par The second class of approaches relies on relaxing the integer constraints in the original CPOP/IPP to form an LPP, which is then solved iteratively by-- invoking an LP solver and generating new pairings using CG; and integerizing the resulting LPP solution. In any iteration of the LPP-solutioning (referred to as an \textit{LPP iteration}), an LP solver (based on either a \textit{simplex} method or an \textit{interior-point} method\footnote{The class of interior-point methods was first introduced by \cite{karmarkar1984new}. In that, a polynomial-time algorithm, called \textit{Karmarkar's algorithm}, was proposed, which, in contrary to simplex method, searches for the best solution by traversing the interior of the feasible region of the search space.}) is invoked on the input pairing set to find the LPP solution and its corresponding dual information (shadow price corresponding to each flight-coverage constraint), which are then utilized to generate new pairings as part of the pricing sub-problem, promising the corresponding cost-improvements. For the first LPP iteration, any set of pairings covering all the flights becomes the input to the LP solver, and for any subsequent LPP iteration, the current LPP solution and the set of new pairings (from the pricing sub-problem) constitute the new input. For more details on how new pairings are generated under the pricing sub-problem in the CG technique, interested readers are referred to \cite{vance1997heuristic, lubbecke2005selected}. As cited in \cite{zeren2016novel}, the CG technique has several limitations, out of which the prominent ones are-- \textit{heading-in} effect (poor dual information in initial LPP iteration leads to generation of irrelevant columns), \textit{bang-bang} effect (dual variables oscillate from one extreme point to another, leading to poor or slower convergence), and \textit{tailing-off} effect (the cost-improvements in the later LPP iterations taper-off). While, different stabilization techniques are available for CG in the literature \cite{du1999stabilized, lubbecke2010column}, the use of interior point methods is gaining prominence. \cite{anbil1991recent} presented the advancements at the American Airlines, and enhanced the approach proposed by \cite{gershkoff1989optimizing} (discussed above), by leveraging the knowledge of dual variables to screen-out/price-out the pairings from the enumerated set at each iteration, enabling it to solve larger sub-problems (up to 25 flights and 100,000 pairings). As an outcome of a collaboration between IBM and American Airlines, \cite{anbil1992global} proposed an iterative global solution approach (though falling short of global optimization) in which an exhaustive set of pairings ($\approx$5.5 million) is enumerated a priori. Several thousands of these pairings are used to initialize the iterative procedure, and in each of its iterations, a sub-problem is solved to obtain optimal dual variables, which are then used to price-out all 5.5 million pairings to select a sufficiently-sized set of good pairings ($\approx$5,000 pairings). For integerization of the LPP solution, the literature points to two prominent strategies. The first strategy is based on utilizing either a branch-and-bound, or a branch-and-cut algorithm. The other strategy utilizes some special ``connection-fixing'' heuristics either solely for integerization \citep{anbil1992global, marsten1994crew}, or during the iterations of the LPP-solutioning \citep{zeren2016novel} to boost the performance of the subsequent MIP solver (in some cases, may even get integer solution without using the MIP solver). These heuristics eliminate some irrelevant pairings by exploiting the knowledge of their linear variables and fixing some specific flight-connections. The limitation in this class of approaches is that even though, a good IPP solution to the original CPOP may exist and the LPP-solutioning leads to a near-optimal LPP solution, the pairings available in it may not fit well together to constitute a good-cost IPP solution.
\par The third class of approaches share a similar solution-architecture as of the preceding class, however, differs in terms of the integerization of the LPP solution. In that, a heuristic branch-and-price framework\footnote{The branch-and-price algorithm was originally proposed by \cite{barnhart1998branch} as an exact algorithm to solve then-known large-scale IPPs, and has been utilized to solve a variety of combinatorial optimization problems in transportation such as \cite{desrosiers1984routing, desrochers1989column, desrochers1992new}.} is adopted, wherein, CG is utilized during the integerization phase too, to generate new legal pairings at nodes of the MIP-search tree. \cite{desrosiers1991breakthrough} is the first instance that solved CPOP using a branch-and-price framework. However, given the inestimable number of legal pairings possible for even medium-scale CPOPs, numerous branch-and-price based heuristic-approaches have been proposed over the last three decades \citep{desaulniers1997crew, vance1997heuristic, anbil1998column, desaulniers2010airline}. Notably, the development of these approaches, being heuristic in nature, require a significant number of algorithmic-design choices to be taken empirically, which may vary with the characteristics of the flight networks being solved for. To name a few such decisions, which branching scheme should be employed (branching on linear variables, branching on flight-connections, or others), should CG be performed on each node of the MIP-search tree, how many CG iterations to be performed each time, etc. Furthermore, the commercial LP and MIP solvers are not much open to modifications, making it difficult for the new researchers to implement a computationally- and time-efficient branch-and-price framework from scratch. For further details of the existing CPOP solution approaches, interested readers are referred to recent survey articles-- \cite{kasirzadeh2017airline, deveci2018survey}.
\par In addition to the above classification of solution approaches, the literature differs on the notion of how the pricing sub-problem is modeled and solved to generate new legal pairings during the LPP iterations. However, the focus of this paper is not on the solution to the pricing sub-problem step, but on the interactions between different modules of a CG-based CPOP solution approach. Hence, for details on the existing work related to the pricing sub-problem step, interested readers are referred to \cite{vance1997heuristic, aggarwal2020novel}.

\subsection{Integer Programming Problem Formulation} \label{sec:model}
As mentioned earlier, a CPOP is intrinsically an IPP, modeled either as a SCP or a SPP. Notably, the SCP formulation provides higher flexibility during its solutioning compared to the SPP formulation by accommodating deadhead flights in the model, possibly resulting in faster convergence \citep{gustafsson1999heuristic}. For a given flight set $\mathcal{F}$ (including $F$ flights) that could be covered in numerous ways by a set of legal pairings $\mathcal{P}$ (including $P$ pairings), the set covering problem is aimed to find a subset of pairings ($\in \mathcal{P}$), say $\mathcal{P}_{IP}^*$, which not only covers each flight ($\in \mathcal{F}$) \textit{at least once}, but does it at a cost \textit{lower} than any alternative subset of pairings in $\mathcal{P}$. In that, while finding $\mathcal{P}_{IP}^*$ ($\subseteq \mathcal{P}$), each pairing $p_j \in \mathcal{P}$ corresponds to a binary variable $x_j$, which represents whether the pairing $p_j$ is included in $\mathcal{P}_{IP}^*$ (marked by $x_j=1$) or not ($x_j=0$). Here, $p_j$ is a $F$-dimensional vector, whose each element, say $a_{ij}$, represents whether the flight $f_i$ is covered by pairing $p_j$ (marked by $a_{ij}=1$) or not ($a_{ij}=0$). In this background, the IPP formulation, as used in this paper, is as follows.
\begin{flalign}
	&\text{Minimize}~ Z_{IP}  = \sum_{j=1}^{P} c_j x_j+ \psi_D \cdot \left(\sum_{i=1}^{F} \left(\sum_{j=1}^{P} a_{ij} x_j - 1\right) \right), \label{eq:obj} &\\
	&\text{subject to} \quad \sum_{j=1}^{P} a_{ij} x_{j} \geq 1,\quad ~~~~~\forall i \in \{1,2,...,F\} \label{eq:coverage}\\
	&\qquad \qquad \quad x_j \in \mathbb{Z}  = \{0,1\},~~~~\forall j \in \{1,2,...,P\} \label{eq:integrality}
\end{flalign}
\begin{flalign}
	\text{where},
	c_j &:~\text{the cost of a legal pairing }p_j,\nonumber &\\
	\psi_D &:~\text{an airline-defined penalty cost against each deadhead in the solution},\nonumber \\
	\quad a_{ij} &=~ 1,~\text{if flight}~f_i~\text{is covered in pairing}~p_j;\text{ else } 0 \nonumber \\
	x_j &=~ 1,~\text{if pairing}~p_j~\text{contributes to Minimum}~Z;\text{ else } 0 \nonumber
\end{flalign}
In the objective function (Equation~\ref{eq:obj}), the first component gives the sum of the individual costs of the pairings selected in the solution, while the other component gives the penalty cost for the deadheads incurred in the solution (note, $(\sum_{j=1}^{P} a_{ij} x_j - 1)$ gives the number of deadheads, corresponding to the flight $f_i$). Notably, in the above formulation, it is assumed that the set of \textit{all} possible legal pairings, namely, $\mathcal{P}$, are available \textit{a priori}, and the task is to determine $\mathcal{P}_{IP}^*$. However, the generation of $\mathcal{P}$ a priori is computationally-intractable for large-scale CPOPs, as mentioned in Section~\ref{sec:relatedwork}. Hence, the solution to the CPOP/IPP is pursued in conjunction with the corresponding LPP (formulation deferred till Section~\ref{sec:LPP}) assisted by the CG technique.

\section{Proposed Airline Crew Pairing Optimization Framework (\textit{AirCROP})} \label{sec:aircrop}
\begin{figure}[pos=htbp,align=\justify]
	\centering \includegraphics[width=0.85\columnwidth, keepaspectratio]{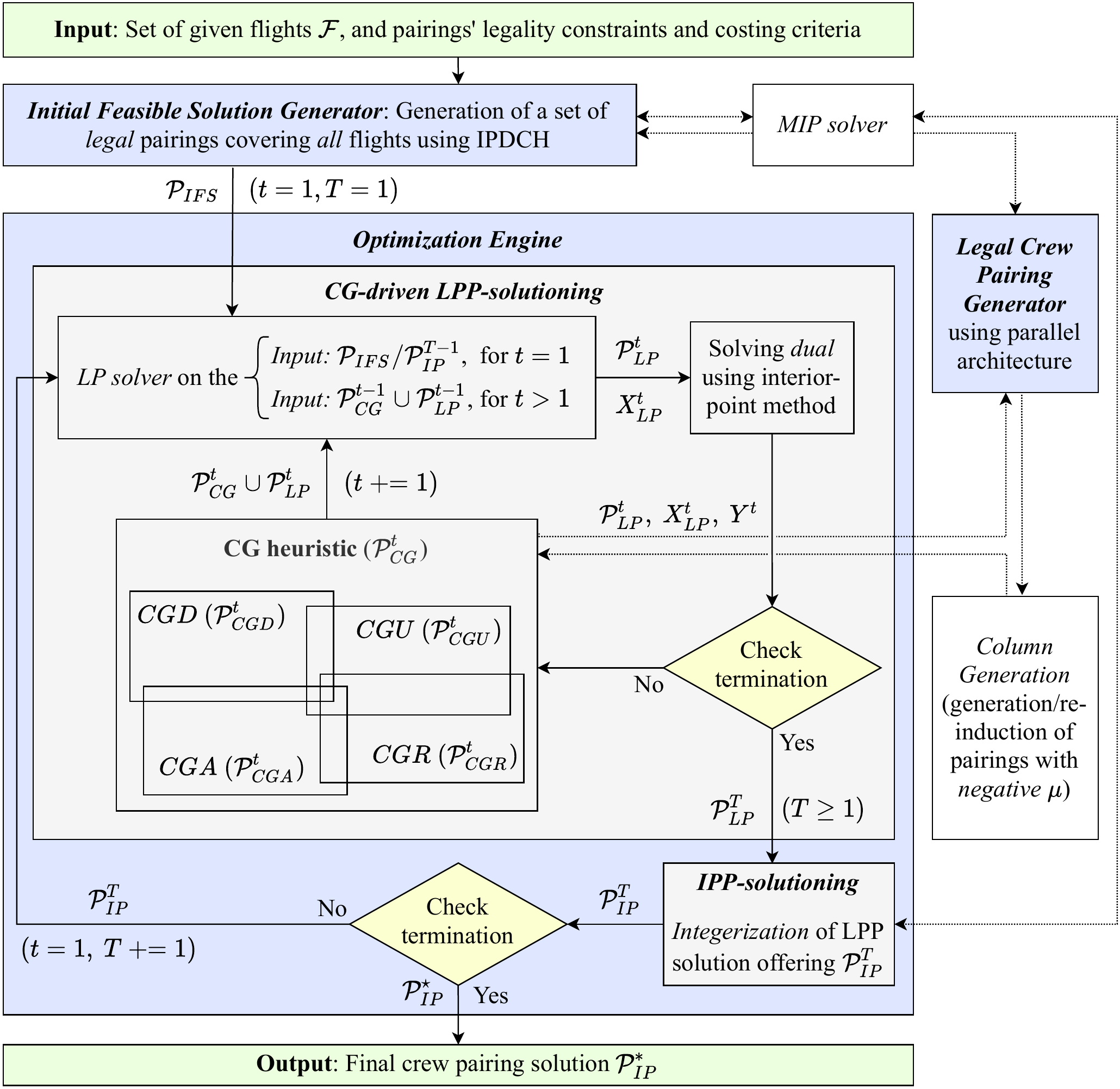}
	\caption{A schematic of $AirCROP$ illustrating the interactions between its constitutive modules-- Legal Crew Pairing Generator, Initial Feasible Solution Generator, Optimization Engine (CG-driven LPP-solutioning interacting with IPP-solutioning). The CG heuristic in LPP-solutioning generates a set of fresh pairings $\mathcal{P}_{CG}^t$ at any LPP iteration $t$ using the following CG strategies: \textit{Deadhead reduction} ($CGD$, generating $\mathcal{P}_{CGD}^t$), \textit{Crew Utilization enhancement} ($CGU$, generating $\mathcal{P}_{CGU}^t$), \textit{Archiving} ($CGA$, generating $\mathcal{P}_{CGA}^t$), and \textit{Random exploration} ($CGR$, generating $\mathcal{P}_{CGR}^t$). The interactions between LPP-solutioning and IPP-solutioning are tracked by the counter $T$.}
	\label{fig:aircrop}
\end{figure}

This section presents the constitutive modules of the proposed optimization framework - $AirCROP$, their working, and their interactions. As per the schematic in Figure~\ref{fig:aircrop}, $AirCROP$ accepts a set of given flights $\mathcal{F}$ along with the pairings' legality constraints and costing criterion as input, and outputs a minimal-cost set of legal pairings $\mathcal{P}_{IP}^\star$, that covers all given flights. This transition from the input to output is enabled by the constitutive modules, namely, the \textit{Legal Crew Pairing Generator}, the \textit{Initial Feasible Solution Generator}, and an \textit{Optimization Engine} in turn enabled by \textit{CG-driven LPP-solutioning} and \textit{IPP-solutioning} submodules and their intermittent interactions. While parts of these modules have been presented elsewhere~\citep{aggarwal2018large, aggarwal2020novel} in isolation, these are being detailed below towards a holistic view on the experimental results presented later.

\subsection{Legal Crew Pairing Generator} \label{sec:LPGP}
This module enables generation of the legal pairings in a time-efficient manner, so they could feed real-time into the other modules - Initial Feasible Solution Generator and the optimization engine. For time-efficiency, it employs a parallel, duty-network based legal pairing generation approach, whose distinctive contributions are  two-folds. Firstly, a crew base centric parallel architecture is adopted considering that several duty- and pairing- constitutive constraints do vary with crew bases. In that, for an input flight set, the legal pairing generation process is decomposed into independent sub-processes (one for each crew base), running in parallel on idle-cores of the central processing unit (CPU). This leads to a significant reduction in the pairing generation time ($\approx$10 folds for a CPOP with 15 crew bases, as demonstrated in \cite{aggarwal2018large}). Secondly, the \textit{set of all possible legal duties} and the corresponding \textit{duty overnight-connection graph} with-respect-to each crew base are enumerated and stored \textit{a priori} the CPOP-solutioning. In a duty overnight-connection graph, a node represents a legal duty, and an edge between any two nodes represents a legal overnight-rest connection between the respective duties. Such a preprocessing ensures that all the connection-city, sit-time, duty, and overnight-rest constraints get naturally satisfied, eliminating the need for their re-evaluation during the generation of legal pairings, and leading to a significant reduction in the legal pairing generation time.
\par The implementation of this module, formalized in Algorithms~\ref{algo:dutyGen}~\&~\ref{algo:pairGen}, is elaborated below. 
\begin{algorithm}[htbp]
	\small
	\DontPrintSemicolon
	\SetKwComment{Comment}{\scriptsize{$\triangleright$\ }}{}
	\KwIn{\nosemic $\mathcal{F}$; $\mathcal{B}$; and constraints: $\mathcal{C}_{connect},~\mathcal{C}_{sit},~\mathcal{C}_{duty}~\&~\mathcal{C}_{night}$\;}
	\KwOut{$\mathcal{D}_{b}$ \& $\mathcal{G}^d_{b}~\forall b \in \mathcal{B}$\;}
	$\mathcal{G}^{f} \gets$ Generate the flight-connection graph by evaluating $\mathcal{C}_{connect}~\&~\mathcal{C}_{sit}$ between each pair of flights $\in \mathcal{F}$\Comment*{$\mathcal{G}^{f} \equiv \left(\mathcal{F},\mathcal{E}^{f}\right)$}
	\For{each crew base $b \in \mathcal{B}$ \Kwpara}{
		\For{each flight $f \in \mathcal{F}$}{
			Push $f$ into an empty $duty$\;
			\If{updated flight-sequence in $duty$ satisfies constraints in $\mathcal{C}_{duty}$}{
				Add $duty$ to $\mathcal{D}_{b}$\;
				\If{$f$ has at least one flight-connection in $\mathcal{G}^{f}$}{
					$\texttt{DFS(}duty, f, \mathcal{G}^{f}, \mathcal{C}_{duty}\texttt{)}$, and add the enumerated duties to $\mathcal{D}_{b}$\;
				}
			}
			Pop out $f$ from $duty$\;
		}
		$\mathcal{G}^{d}_b \gets$ Generate the duty overnight-connection graph by evaluating $\mathcal{C}_{night}$ between each pair of duties $\in \mathcal{D}_b$\;
	}
	\textbf{return} $\mathcal{D}_{b}$ \& $\mathcal{G}^{d}_{b}~\forall b \in \mathcal{B}$\;
	\vspace{5pt}\Comment{DFS($duty,parent,\mathcal{G}^{f},\mathcal{C}_{duty}$)}
	\For{each $child$ of $parent$ in $\mathcal{G}^{f}$}{
		Push $child$ into $duty$\;
		\If{updated flight-sequence in $duty$ satisfies $\mathcal{C}_{duty}$}{
			\textbf{yield} $duty$ to $\mathcal{D}_{b}$\;
			\If{$child$ has at least one connection in $\mathcal{G}^{f}$}{
				\texttt{DFS(}$duty,child,\mathcal{G}^{f},\mathcal{C}_{duty}$\texttt{)}\;
			}
		}
		Pop out $child$ from $duty$\;
	}
	\caption{Procedure for enumeration of legal duties and duty overnight-connection graphs}
	\label{algo:dutyGen}
\end{algorithm}
\begin{algorithm}[htbp]
	\small
	\DontPrintSemicolon
	\SetKwComment{Comment}{\scriptsize{$\triangleright$\ }}{}
	\KwIn{\nosemic $\mathcal{F}_*\text{ or }\mathcal{D}_*;~\mathcal{B};~\mathcal{D}_{b}~\&~\mathcal{G}^{d}_{b}~\forall b \in \mathcal{B}$; and constraints: $\mathcal{C}_{base}~\&~\mathcal{C}_{other}$\;}
	\KwOut{$\mathcal{P}_*$\;}
	\For{each crew base $b \in \mathcal{B}$ \Kwpara}{
		Update $\mathcal{D}_{b}~\&~\mathcal{G}^{d}_{b}$ by removing duties $\notin \mathcal{D}_*$ if $\mathcal{D}_*$ is input, or by removing those duties which cover flights $\notin \mathcal{F}_*$ if $\mathcal{F}_*$ is input\;
		\For{each $duty \in \mathcal{D}_{b}$}{
			\If{departure airport of $duty$ is $b$}{
				Push $duty$ into an empty $pairing$\;
				\If{updated duty-sequence in $pairing$ satisfies $\mathcal{C}_{other}$}{
					\uIf{updated duty-sequence in $pairing$ satisfies $\mathcal{C}_{base}$}{
						Add $pairing$ to $\mathcal{P}_*$\;
					}
					\ElseIf{$duty$ has at least one duty overnight-connection in $\mathcal{G}^{d}_{b}$}{
						$\texttt{DFS(}pairing, duty,\mathcal{G}^{d}_{b},\mathcal{C}_{base} \cup \mathcal{C}_{other}\texttt{)}$, and add enumerated pairings to $\mathcal{P}_*$\;
					}
				}
				Pop out $duty$ from $pairing$\;
			}
		}
	}
	\textbf{return} $\mathcal{P}_*$\;
	\vspace{5pt}\Comment{DFS($pairing,parent,\mathcal{G}^{d}_{b},\mathcal{C}_{base} \cup \mathcal{C}_{other}$)}
	\For{each $child$ of $parent$ in $\mathcal{G}^{d}_b$}{
		Push $child$ into $pairing$\;
		\If{updated duty-sequence in $pairing$ satisfies $\mathcal{C}_{other}$}{
			\uIf{updated duty-sequence in $pairing$ satisfies $\mathcal{C}_{base}$}{
				\textbf{yield} $pairing$ to $\mathcal{P}_*$\;
			}
			\ElseIf{$child$ has at least one duty overnight-connection in $\mathcal{G}^{d}_{b}$}{
				$\texttt{DFS(}pairing, child,\mathcal{G}^{d}_{b},\mathcal{C}_{base} \cup \mathcal{C}_{other}\texttt{)}$\;
			}
		}
		Pop out $child$ from $pairing$\;
	}
	\caption{Procedure for enumeration of legal pairings from an input flight set $\mathcal{F}_*\text{ or a duty set }\mathcal{D}_*$}
	\label{algo:pairGen}
\end{algorithm}
For solving any CPOP, the foremost step of the $AirCROP$ is to preprocess the entire duty-connection network-- set of legal duties $\mathcal{D}_{b}$ and duty overnight-connection graph $\mathcal{G}^d_{b} \left(\equiv \left(\mathcal{D}_{b},~\mathcal{E}^d_{b}\right)\right)$ for each crew base $b$ in the given set of crew bases $\mathcal{B}$, where $\mathcal{E}^d_{b}$ is the set of legal overnight-rest connections between duty-pairs $\in \mathcal{D}_b$. The procedure for the above preprocessing is presented in Algorithm~\ref{algo:dutyGen}. In that, the first step is the generation of a flight-connection graph (denoted by $\mathcal{G}^{f}$) by evaluating the legality of connection-city ($\mathcal{C}_{connect}$) and sit-time ($\mathcal{C}_{sit}$) constraints between every flight-pair in the given flight schedule $\mathcal{F}$ (line 1). Here, in $\mathcal{G}^{f}~\left(\equiv \left(\mathcal{F},\mathcal{E}^{f}\right)\right)$, $\mathcal{F}$ is the set of nodes (flights) and $\mathcal{E}^{f}$ is the set of edges (legal flight connections). Subsequently, $\mathcal{G}^f$ is used for legal duty enumeration, by decomposing the process into independent sub-processes, one for each crew base $b \in \mathcal{B}$, and executing them in parallel (lines 2-12). In each of these sub-processes, enumeration of legal duties, starting from each flight $f \in \mathcal{F}$, is explored. In that:
\begin{itemize}
	\item flight $f$ is added to an empty candidate duty stack, given by $duty$ (line 4).
	\item the flight-sequence in $duty$ is checked for satisfaction of duty constraints $\mathcal{C}_{duty}$, and if satisfied, $duty$ is added to the desired legal duty set $\mathcal{D}_{b}$ (lines 5-6). Notably, if $f$ has at least one connection with another flight in $\mathcal{G}^{f}$, and if the duty constraints permit, then more flights could be accommodated in $duty$, leading to enumeration of other legal duties (lines 7-9).
	\item a Depth-first Search (DFS) algorithm \citep{tarjan1972depth} is adapted, which is called recursively to enumerate legal duties, starting from a parent flight node ($parent$), by exploring its all successive paths in $\mathcal{G}^{f}$ in a depth-first manner (lines 16-25). In each recursion, a child flight node ($child$) is pushed into $duty$, the updated flight-sequence is checked for satisfaction of $\mathcal{C}_{duty}$, and if satisfied, $duty$ is yielded to $\mathcal{D}_b$, followed by another recursion of \texttt{DFS()} with $child$ as the new $parent$. In this way, all legal duties, starting from flight $f$, are enumerated.
\end{itemize}
Subsequently, $f$ is popped out from $duty$, and duty enumeration using other flights in $\mathcal{F}$ is explored (lines 3 \& 11). The resulting set $\mathcal{D}_{b}$ is then used to generate the duty overnight-connection graph $\mathcal{G}^{d}_{b}$), by evaluating the legality of connection-city ($\mathcal{C}_{connect}$) and overnight-rest ($\mathcal{C}_{night}$) constraints between every duty-pair $\in \mathcal{D}_{b}$ (line 13). Here, in $\mathcal{G}^{d}_b~\left(\equiv \left(\mathcal{D}_b,\mathcal{E}^{d}_b\right)\right)$, $\mathcal{D}_b$ is the set of nodes (legal duties), and $\mathcal{E}^{d}_b$ is the set of edges (legal overnight-rest connections).
\par The preprocessed sets of legal duties and the corresponding duty overnight-connection graphs are utilized to enumerate legal pairings for any input flight set (say $\mathcal{F}_*$) or a duty set (say $\mathcal{D}_*$), when required in real-time in other modules of the $AirCROP$. Its procedure, formalized in Algorithm~\ref{algo:pairGen}, is elaborated below. For legal pairing enumeration, the same crew base driven parallel architecture is utilized in which the process is decomposed into independent sub-processes, one for each crew base $b \in \mathcal{B}$, running in parallel on idle-cores of the CPU (line 1). In each of these sub-processes, the first step is to update $\mathcal{D}_{b}$ and $\mathcal{G}^{d}_{b}$, by removing duties $\notin \mathcal{D}_*$ if $\mathcal{D}_*$ is input, or those duties that cover flights $\notin \mathcal{F}_*$ if $\mathcal{F}_*$ is input (line 2). Subsequently, the enumeration of legal pairings, starting from each duty ($duty$) $\in \mathcal{D}_{b}$, is explored (line 3). In that:
\begin{itemize}
	\item the $duty$ is pushed into an empty candidate pairing stack, given by $pairing$, only if the departure airport of $duty$ is same as the crew base $b$ (lines 4-5).
	\item the $pairing$ is checked for satisfaction of pairing constraints $\mathcal{C}_{other}$, and if satisfied, $pairing$ is further checked for satisfaction of end-city constraint $\mathcal{C}_{base}$, which ensures that the arrival airport of the $pairing$'s last duty is same as the crew base $b$.
	\begin{itemize}
		\item If $pairing$ satisfies $\mathcal{C}_{base}$, it is classified as \textit{legal}, and is added to the desired pairing set $\mathcal{P}_*$ (lines 7-8).
		\item If $pairing$ does not satisfy $\mathcal{C}_{base}$, it is not complete, and more duties are required to be covered in it to complete the legal duty-sequence. This is only possible if $duty$ has at least one overnight-rest connection in $\mathcal{G}^{d}_{b}$. And if it does, the \texttt{DFS()} sub-routine, similar to the one used in legal duty enumeration, is called recursively to enumerate legal pairings, starting from a parent duty node ($parent$), by exploring its all successive paths in $\mathcal{G}^{d}_b$ in a depth-first manner (lines 18-28). In each recursion:
		\begin{itemize}[label=$\circ$]
			\item a child duty node ($child$) is pushed into the $pairing$ (line 19).
			\item the updated duty-sequence in $pairing$ is checked for satisfaction of first $\mathcal{C}_{other}$ and then $\mathcal{C}_{base}$ (lines 20-21).
			\item if it satisfies both constraints, then $pairing$ is complete (legal), and is yielded to the desired pairing set $\mathcal{P}_*$ (line 22).
			\item if it satisfies $\mathcal{C}_{other}$ but not $\mathcal{C}_{base}$, then another recursion of \texttt{DFS()} with $child$ as new $parent$ is called, only if $child$ has at least one duty overnight-rest connection in $\mathcal{G}^{d}_b$ (lines 23-25).
		\end{itemize}
	\end{itemize}
\end{itemize}
In the above way, all legal pairings, starting from $duty$, are enumerated using the \texttt{DFS()} sub-routine. Subsequently, $duty$ is popped out of $pairing$ (line 13), and the legal pairing enumeration using other duties $\in \mathcal{D}_{b}$ is explored (line 3). Once, all the sub-processes are complete, the desired pairing set $\mathcal{P}_*$ is returned (line 17).

\subsection{Initial Feasible Solution Generator} \label{sec:IFS}
An initial feasible solution (IFS) is \textit{any} set of pairings, covering \textit{all} flights in the given flight schedule, which is used to initialize a CPOP solution approach. For large-scale CPOPs, generation of an IFS standalone is a computationally-challenging task. This module is designed to generate a reasonably-sized IFS in a time-efficient manner for large and complex flight networks, which is then used to initialize the Optimization Engine of $AirCROP$. For this, it employs a novel \textit{Integer Programming based Divide-and-cover Heuristic} (IPDCH), which relies on: (a) a \textit{divide-and-cover} strategy to decompose the input flight schedule into sufficiently-small flight subsets, and (b) \textit{integer programming} to find a lowest-cost pairing set, covering the maximum possible flights for each of the decomposed flight subsets.
\par The procedure of the proposed IPDCH, formalized in Algorithm~\ref{algo:IPDCH}, is elaborated below.
\begin{algorithm}[htbp]
	\small
	\DontPrintSemicolon
	\SetKwComment{Comment}{\scriptsize{$\triangleright$\ }}{}
	\KwIn {$\mathcal{F},~K,~\texttt{Pairing\_Gen()}$\;}
	\KwOut{$\mathcal{P}_{IFS}$\;}
	\While{all flights $\in \mathcal{F}$ are not covered in $\mathcal{P}_{IFS}$}{
		$\mathcal{F}_{K} \gets $ Select $K$ random flights from $\mathcal{F}$ without replacement \Comment*{$K < F$}
		$\mathcal{P}_{K} \gets~\texttt{Pairing\_Gen(}\mathcal{F}_K\texttt{)}$\;
		$\mathcal{F}_{K'} \gets $ Flights covered in $\mathcal{P}_{K}$ \Comment*{$K' \leq K$}
		Add remaining flights $\left(\mathcal{F}_{K} \backslash \mathcal{F}_{K'}\right)$ back to $\mathcal{F}_{K}$\;
		Formulate the IPP using flights in $\mathcal{F}_{K'}$ and pairings in $\mathcal{P}_{K}$\;
		$\mathcal{P}_{IP} \gets$ Solve the IPP using an MIP solver, and select pairings corresponding to non-zero variables\;
		Add pairings from $\mathcal{P}_{IP}$ to $\mathcal{P}_{IFS}$\;
		Replace flights in $\mathcal{F}$ if it becomes empty\;		
	}
	\textbf{return} $\mathcal{P}_{IFS} $\;
	\caption{Procedure for IFS generation using the proposed IPDCH}
	\label{algo:IPDCH}
\end{algorithm}
Being an iterative heuristic, IPDCH terminates when all flights in the input set are covered by pairings in the desired IFS, notated as $\mathcal{P}_{IFS}$ (lines 1). The input to the heuristic involves the given flight schedule $\mathcal{F}$ (with $F$ number of flights), the pairing generation sub-routine \texttt{Pairing\_Gen()} (presented in Section~\ref{sec:LPGP}), and a pre-defined \textit{decomposition parameter} $K$, which regulates the number of flights to be selected from $\mathcal{F}$ in each IPDCH-iteration. The setting of $K$ largely depends upon the available computational resources, and the characteristics of the input flight dataset (as highlighted in Section~\ref{sec:computeInit}). In each IPDCH-iteration, first a flight subset, say $\mathcal{F}_{K}$ $\left(K<F\right)$, is formed by randomly selecting $K$ number of flights from $\mathcal{F}$ without replacement (line 2). Subsequently, $\mathcal{F}_{K}$ is fed as input to the \texttt{Pairing\_Gen()} sub-routine to enumerate the set of all possible legal pairings, say $\mathcal{P}_{K}$ (line 3). Notably, all flights in $\mathcal{F}_{K}$ may not get covered by pairings in $\mathcal{P}_{K}$, as random selection of flights does not guarantee legal connections for all selected flights. Let $\mathcal{F}_{K'}$ $\left(K' \leq K\right)$ be the set of flights covered in $\mathcal{P}_{K}$ (line 4). The remaining flights, given by $\mathcal{F}_{K} \backslash \mathcal{F}_{K'}$, are added back to $\mathcal{F}$ (line 5). Subsequently, $\mathcal{F}_{K'}$ and $\mathcal{P}_{K}$ are used to formulate the corresponding IPP (line 6), which is then solved using a commercial off-the-shelf MIP solver to find the optimal IPP solution, say $\mathcal{P}_{IP}$, constituted by pairings corresponding to only non-zero variables (line 7). The pairings in $\mathcal{P}_{IP}$ are then added to the desired set $\mathcal{P}_{IFS}$ (line 8). Lastly, the flights in $\mathcal{F}$ are replaced if it becomes empty (line 9). As soon as $\mathcal{P}_{IFS}$ covers all the required flights, IPDCH is terminated, and $\mathcal{P}_{IFS}$ is passed over to the Optimization Engine for its initialization.

\subsection{Optimization Engine: Interactions between CG-driven LPP-solutioning and IPP-solutioning} \label{sec:optEng}
The search for minimal cost, full flight-coverage CPOP solution is enabled by an optimization engine. It tackles the underlying LPP and IPP through intermittent interactions of two submodules, namely, CG-driven LPP-solutioning and IPP-solutioning, tracked by a counter $T$. These submodules are presented below.

\subsubsection{CG-driven LPP-solutioning} \label{sec:LPP}
As illustrated in Figure~\ref{fig:aircrop}, this submodule entails several iterations  (each referred to as an \textit{LPP iteration}, and is tracked by $t$) in each of which: (a) an LP solver is invoked on the input pairing set, leading to the current LPP solution $\mathcal{P}_{LP}^t$, (b) the corresponding \textit{dual} od the LPP is formulated using $\mathcal{P}_{LP}^t$, which is then solved to fetch dual variables (given by vector $Y^t$), and (c) a fresh set of pairings $\mathcal{P}_{CG}^t$, that promises associated cost-improvement, is generated using a domain-knowledge driven CG heuristic. For the first LPP iteration ($t=1$), the input to the LP solver is either $\mathcal{P}_{IFS}$ if $T=1$, or $\mathcal{P}_{IP}^{T-1}$ if $T>1$. For any subsequent LPP iteration ($t>1$), the input comprises of the current $\mathcal{P}_{CG}^t$ and $\mathcal{P}_{LP}^{t}$.
In this background, each of these LPP iterations are implemented in the following three phases\footnote{For ease of reference, the notations introduced in these phases are kept independent of the LPP iteration counter $t$. However, these notations are super-scripted by $t$ in the corresponding discussions and pseudocodes with reference to a particular LPP iteration.}:
\begin{itemize}
	\item In the first phase, a \textit{primal} of the LPP (Equations~\ref{eq:obj2}~to~\ref{eq:integrality2}) is formulated from the input pairing set, and is solved using an interior-point method based commercial off-the-shelf LP solver \citep{gurobi}. In the resulting LPP solution, a \textit{primal variable} $x_j$, varying from $0$ to $1$, is assigned to each pairing $p_j$ in the input pairing set. These $x_j$s together constitute the \textit{primal vector}, notated as $X~\left(=[x_1~x_2~x_3~...~x_P]^{\mathsf{T}}\right)$. The set of $x_j$s with non-zero values ($x_j \neq 0$) and the set of corresponding pairings are notated as $X_{LP}$ and $\mathcal{P}_{LP}$, respectively.
	\begin{flalign}
		&\text{Minimize}~ Z_{LP}^p = \sum_{j=1}^{P} c_j x_j+ \psi_D \cdot \left(\sum_{i=1}^{F} \left(\sum_{j=1}^{P} a_{ij} x_j - 1\right) \right)=\sum_{j=1}^{P} \left(c_j + \psi_D \cdot \sum_{i=1}^{F} a_{ij} \right) x_j - F \cdot \psi_D, &\label{eq:obj2} \\
		&\text{subject to} \quad \sum_{j=1}^{P} a_{ij} x_{j} \geq 1,\qquad \quad \forall i \in \{1,2,...,F\} \label{eq:coverage2}\\
		&\qquad \qquad \quad x_j \in \mathbb{R} = [0,1],\qquad \forall j \in \{1,2,...,P\} \label{eq:integrality2}
	\end{flalign}
	It is to be noted that the minimization of $Z_{LP}^p$ will always lead to a solution with all primal variables $x_j \leq 1$, even without explicitly involving the corresponding constraint-- Equation~\ref{eq:integrality2} \citep{vazirani2013approximation}. Hence, the contribution of each pairing in the LPP solution, given by its $x_j$, could be effectively treated as $x_j \in \mathbb{R}_{\geq 0}$ instead of Equation~\ref{eq:integrality2}.
	\item In the second phase, dual variables are extracted from the current LPP solution. For this, the \textit{dual} of the LPP (Equations~\ref{eq:dualobj}~to~\ref{eq:dualvariables}) is formulated using the pairing set $\mathcal{P}_{LP}$, and is solved using an interior-point method \citep{andersen2000mosek} based non-commercial LP solver \citep{scipy}, to fetch the optimal dual solution. In that, a \textit{dual variable} $y_i$ represents a shadow price corresponding to an $i^{th}$ flight-coverage constraint in the primal. The \textit{optimal dual vector}, constituted by all $y_i$s in the optimal dual solution, is notated as $Y~\left(=[y_1~y_2~y_3~...~y_F]^{\mathsf{T}}\right)$, whose dimension is equal to $F$.
	\begin{flalign}
		&\text{Maximize}~ Z_{LP}^d = \sum_{i=1}^{F} y_i - F \cdot \psi_D, \label{eq:dualobj} &\\
		&\text{subject to} \quad \sum_{i=1}^{F} a_{ij} y_i \leq \left(c_j + \psi_D \cdot \sum_{i=1}^{F} a_{ij}\right),~~~~\forall j \in \{1,2,...,P_{LP}\} \label{eq:dualconstraints}\\
		& \qquad \qquad \qquad \quad~~y_i \in \mathbb{R} \geq 0,\qquad \qquad \qquad~~~~\forall i \in \{1,2,...,F\} \label{eq:dualvariables}\\
		&\text{where}, \qquad P_{LP} :~\text{is the number of pairings in the set}~  \mathcal{P}_{LP}  \nonumber \\
		&\qquad \qquad \quad~~~y_i :~\text{dual variable, corresponding to an $i^{th}$ flight-coverage constraint}, \nonumber 
	\end{flalign}
	Notably, in a conventional approach, the optimal $Y$ is directly computed from the optimal basis of the primal solution (obtained in the first phase), using the principles of \textit{duality theory}, particularly the \textit{theorem of complementary slackness} \citep{bertsimas1997introduction}, without explicitly solving the corresponding dual. However, in the second phase, solving the dual explicitly using the interior-point method \citep{andersen2000mosek}, in a sense, helps in stabilizing the oscillating behavior of dual variables over the successive LPP iterations (bang-bang effect, as discussed in Section~\ref{sec:relatedwork}). Moreover, this interior-point method is available via only a non-commercial LP solver \citep{scipy}, and to ensure a time-efficient search, the above dual is formulated using the pairings $\in \mathcal{P}_{LP}$, instead of pairings from the large-sized input pairing set.
	\item In the last phase, the availability of \textit{dual variables} from the second phase paves the way for solution to the pricing sub-problem. It is aimed to generate those legal pairings (non-basic), which if included as part of the input to the next LPP iteration, promise a better-cost (at least a similar-cost) LPP solution compared to the current solution. Such non-basic pairings are identified using a \textit{reduced cost} metric, given by $\mu_j$ (Equation~\ref{eq:redCost}), which if negative (as CPOP is a minimization problem) indicates the potential in the pairing to further reduce the cost of the current LPP solution $Z_{LP}^p$, when included in the current basis \citep{bertsimas1997introduction}. Moreover, the potential of such a pairing to further reduce the current $Z_{LP}^p$, is in proportion to the magnitude of its $\mu_j$ value.
	\begin{flalign}
		&\mu_j = c_j - \mu d_j,~\text{where,}~\mu d_j = \sum_{i=1}^{F} \left(a_{ij} \cdot y_i\right)~= \sum_{f_i \in p_j} y_i ~~(~\text{represents the dual cost component of}~\mu_j) & \label{eq:redCost}
	\end{flalign}
	As mentioned in Section~\ref{sec:relatedwork}, the standard CG practices generate a complete pricing network and solves it as a resource-constrained shortest-path optimization problem, to identify only the pairing(s) with negative reduced cost(s). However, generation of a complete pricing network for CPOPs with large-scale and complex flight networks is computationally-intractable. To overcome this challenge, a \textit{domain-knowledge driven CG heuristic} \citep{aggarwal2020novel} is employed here to generate a set of promising pairings (of pre-defined size, criterion for which is discussed in Section~\ref{sec:parameter-setting}). Notably, the merit of this CG heuristic lies in the fact that from within the larger pool of pairings with negative $\mu_j$, besides selecting pairings randomly, it also selects pairings in a guided manner. In that, the selection of such pairings is guided by \textit{optimal solution features} at a \textit{set level} and an individual \textit{pairing level}, and \textit{re-utilization of the past computational efforts}. These optimal solution features are related to	the \textit{minimization of deadheads} and \textit{maximization of the crew utilization}, respectively. In essence, while the standard CG practices present equal opportunity for any pairing with a negative $\mu_j$ to qualify as an input for the next LPP iteration, this CG heuristic, besides ensuring that the pairings have negative $\mu_j$, prioritizes some pairings over the others via its two-pronged strategy-- \textit{exploration of the new pairings' space} and \textit{re-utilization of pairings from the past LPP iterations}. In that:
	\begin{itemize}
		\item the \textit{exploration of the new pairings' space} is guided by three CG strategies, which are elaborated below.
		\begin{itemize}[label=$\circ$]
			\item \textit{Deadhead Reduction strategy} ($CGD$): this strategy prioritizes a set of legal pairings that is characterized by \textit{low deadheads}, a feature which domain knowledge recommends for optimality at a \textit{set level}. To exploit this optimality feature, $CGD$ generates a new paring set $\mathcal{P}_{CGD}$, which not only provides an alternative way to cover the flights involved in a subset of the current $\mathcal{P}_{LP}$, but also ensures that some of these flights get covered with zero deadheads. It promises propagation of the zero deadhead feature over successive LPP iterations, as: (a) $\mathcal{P}_{CGD}$ alongside the current $\mathcal{P}_{LP}$ forms a part of the input for the next LPP iteration; (b) $\mathcal{P}_{CGD}$ provides a scope for better coverage (zero deadhead) of some flights, compared to the current $\mathcal{P}_{LP}$; and (c) $\mathcal{P}_{CGD}$ may focus on zero deadhead coverage for different flights in different LPP iterations.
			\item \textit{Crew Utilization enhancement strategy} ($CGU$): this strategy prioritizes a set of legal pairings each member of which is characterized by \textit{high crew utilization}, a feature which domain knowledge recommends for optimality at an individual \textit{pairing level}. To exploit this optimality feature, $CGU$: (a) introduces a new measure, namely, \textit{crew utilization ratio}, given by $\gamma_j$ (Equation~\ref{eq:crewUtilityRatio}), to quantify the degree of crew utilization in a pairing $p_j$ at any instant; (b) identifies pairings from the current $\mathcal{P}_{LP}$, which are characterized by high dual cost component ($\mu d_j$, Equation~\ref{eq:redCost}), reflecting in turn on those constitutive flights that have high value of dual variables $y_i$, and hence, on the potential of these flights to generate new pairings with more negative $\mu_j$; and (c) utilizes these flights to generate promising pairings from which only the ones with high $\gamma_j$ are picked to constitute the new pairing set $\mathcal{P}_{CGU}$.
			\begin{flalign}
				\gamma_j = \frac{1}{\text{Number of duties in }p_j} \cdot \sum_{d \in p_j} \frac{\text{Working hours in duty} ~d}{\text{Permissible hours of duty } d}  \label{eq:crewUtilityRatio}
			\end{flalign}
			In doing so, $CGD$ promises propagation of the higher crew utilization ratio over successive LPP iterations, given that in each LPP iteration, $\mathcal{P}_{CGU}$ alongside the current $\mathcal{P}_{LP}$ forms a part of the input for the next LPP iteration. 
			\item \textit{Random exploration strategy} ($CGR$): this strategy, unlike $CGU$ and $CGD$ which are guided by optimal solution features, pursues random and unbiased exploration of the new pairings' space, independent of the current LPP solution. It involves generation of new pairings for a random selected set of legal duties from which only the pairings with negative reduced cost are selected to constitute the new pairing set $\mathcal{P}_{CGR}$. Here, a random set of legal duties is used instead of a random set of flights, as the former has a higher probability of generating legal pairings, given that a majority of pairing legality constraints get satisfied with the preprocessing of legal duties. 
		\end{itemize}
		\item the \textit{re-utilization of pairings from the past LPP iterations} is guided by an \textit{Archiving strategy} ($CGA$), that prioritizes a set of legal pairings comprising of those flight-pairs, which as per the existing LPP solution, bear better potential for improvement in the objective function. Such a pairing set, originating from the \textit{flight-pair level} information, is extracted from an \textit{archive} (denoted by $\mathcal{A}$) of the previously generated pairings. In doing so, $CGA$ facilitates re-utilization of the past computational efforts, by providing an opportunity for a previously generated pairing to be re-inducted in the current pairing pool. For this, $CGA$:
		\begin{itemize}[label=$\circ$]
			\item updates the archive $\mathcal{A}$ in each LPP iteration such that any pairing is stored/retrieved with reference to a unique index $(f_m,f_n)$ reserved for any legal flight-pair in that pairing.
			\item introduces a new measure, namely, \textit{reduced cost estimator}, given by $\eta_{mn}$ (Equation~\ref{eq:redCostEst}), for a flight-pair $(f_m,f_n)$ in $\mathcal{A}$. In each LPP iteration, this estimator is computed for all the flight-pairs present in $\mathcal{A}$, by fetching $f_m$, $f_n$, $y_m$ and $y_n$.
			\begin{flalign}
				\eta_{mn} &= \texttt{flying\_cost($f_m$)} +  \texttt{flying\_cost($f_n$)} - y_m - y_n = \sum_{i \in \{m,n\}} \left(\texttt{flying\_cost($f_i$)} - y_i \right) \label{eq:redCostEst}
			\end{flalign}
			Notably, this formulation is analogous to Equation~\ref{eq:redCost}, just that instead of the complete cost of a pairing, only the flying costs corresponding to the flights in a legal flight-pair are accounted for. Given this, $\eta_{mn}$ may be seen as an indicator of $\mu_j$ \textit{at the flight-pair level}.
			\item recognizes that towards further improvement in the current LPP solution, it may be prudent to include as a part of the input for the next LPP iteration-- the new pairing set $\mathcal{P}_{CGA}$, constituted by preferentially picking pairings from $\mathcal{A}$, that cover flight-pairs with lower $\eta_{mn}$ value.
		\end{itemize}
		In doing so, $CGA$ pursues the goal of continual improvement in the objective function, while relying on the \textit{flight-pair} level information embedded in the LPP solution of current LPP iteration, and re-utilizing the computational efforts spent till that LPP iteration.
	\end{itemize}
	For further details and associated nitty-gritty of the above domain-knowledge driven CG heuristic, interested readers are referred to the authors' previous work-- \cite{aggarwal2020novel}. Once this CG heuristic generates a set of promising pairings $\mathcal{P}_{CG}$ of pre-defined size, it is merged with the current $\mathcal{P}_{LP}$, and fed as the input to the next LPP iteration ($t \pluseq 1$).
\end{itemize}
These LPP iterations are repeated until the cost-improvements over a pre-specified number of successive LPP iterations falls below a pre-specified cost-threshold (settings given in Section~\ref{sec:parameter-setting}).
In this submodule, these LPP iterations are repeated, until its termination criterion is not met. In that, the cost-improvement over LPP iterations is observed, and if it falls below a pre-specified cost-threshold, say $Th_{cost}$, over a pre-specified number of successive LPP iterations, say $Th_t$, then it is terminated. The settings of these pre-specified limits-- $Th_{cost}$ and $Th_t$, are highlighted in Section~\ref{sec:parameter-setting}. After termination, the final LPP solution $\mathcal{P}_{LP}^T$ is then passed over to the IPP-solutioning submodule for its integerization.

\subsubsection{IPP-solutioning} \label{sec:IPP}
This submodule receives as input, the LPP solution $\mathcal{P}^T_{LP}$, and aims to find therein a full-coverage integer solution, notated as $\mathcal{P}^T_{IP}$. Towards it, an IPP (Equations~\ref{eq:obj}~to~\ref{eq:integrality}) is formulated using $\mathcal{P}^T_{LP}$ and $\mathcal{F}$, and solved using a branch-and-cut algorithm based off-the-shelf commercial MIP solver \citep{gurobi}. At each node of the MIP-search tree, this solver maintains a valid lower bound (cost of the LPP solution) and a best upper bound (cost of the IPP solution), and it self-terminates if the gap between these two bounds becomes zero, or all branches in the MIP-search tree have been explored. Considering that the MIP-search for large-scale CPOPs is extremely time-consuming, a pre-defined time limit, notated as $Th_{ipt}$ (setting highlighted in Section~\ref{sec:parameter-setting}), is used to terminate this MIP solver, if it does not terminate by itself a priori. Once the $\mathcal{P}^T_{IP}$ is obtained, it is passed back to the previous submodule for the next LPP-IPP interaction ($T \pluseq 1$), only if the termination criterion of the Optimization Engine is not satisfied.\\\\
%
\noindent \textbf{Overarching Optimization Engine}\\
In the wake of the above, the procedure of the overarching Optimization Engine, formalized in Algorithm~\ref{algo:optEng}, is elaborated below. 
\begin{algorithm}[htbp]
	\small
	\DontPrintSemicolon
	\SetKwComment{Comment}{\scriptsize{$\triangleright$\ }}{}
	\KwIn {$\mathcal{F},~\mathcal{P}_{IFS},~Th_{cost},~Th_{t},~Th_{ipt},~\texttt{Pairing\_Gen()},~\texttt{CGD()},~\texttt{CGU()},~\texttt{CGR()},~\texttt{CGA()}$\;}
	\KwOut{$\mathcal{P}^\star_{IP}$\;}
	\nosemic $T \gets 1$\;
	\While{termination criterion of Optimization Engine is not met}{
		\vspace{3pt}\Comment*[f]{CG-driven LPP-solutioning:}\;
		$t \gets 1$\;
		\While{termiantion criterion of CG-driven LPP-solutioning is not met}{
			\uIf{$t=1$ and $T=1$}{
				Formulate the \textit{primal} of the LPP using $\mathcal{P}_{IFS}$ and $\mathcal{F}$\;
			}
			\uElseIf{$t=1$ and $T>1$}{
				Formulate the \textit{primal} of the LPP using $\mathcal{P}^{T-1}_{IP}$ and $\mathcal{F}$\;
			}
			\Else{
				Formulate the \textit{primal} of the LPP using $\mathcal{P}^{t-1}_{CG} \cup \mathcal{P}^{t-1}_{LP}$ and $\mathcal{F}$\;
			}
			$\mathcal{P}^{t}_{LP},~X^t_{LP} \gets$ Solve the \textit{primal} using the interior-point method based LP solver\;
			\vspace{3pt}\Comment*[f]{Termination of the CG-driven LPP-solutioning:}\;
			\If{cost-improvements $\leq Th_{cost}$ over last $Th_t$ number of successive LPP iterations}{
				$\mathcal{P}^T_{LP} \gets \mathcal{P}^{t}_{LP}$\;
				\textbf{Break}\;
			}
			Formulate the \textit{dual} of the LPP using $\mathcal{F}$ and $\mathcal{P}^t_{LP}$\;
			$Y^t \gets $ Solve the dual using the interior-point method based LP solver\;
			\vspace{3pt}\Comment*[f]{Solution to pricing sub-problem using the CG heuristic:}\;
			$\mathcal{P}^{t}_{CGD} \gets \texttt{CGD($\mathcal{P}^{t}_{LP},X^t_{LP},Y^t,\ldots$)}$\;
			$\mathcal{P}^{t}_{CGU} \gets \texttt{CGU($\mathcal{P}^{t}_{LP},X^t_{LP},Y^t,\ldots$)}$\;
			$\mathcal{P}^{t}_{CGR} \gets \texttt{CGR($Y^t,\ldots$)}$\;
			$\mathcal{P}^{t}_{CGA} \gets \texttt{CGA($\mathcal{P}^{t}_{LP},X^t_{LP},Y^t,\ldots$)}$\;
			$\mathcal{P}^{t}_{CG} \gets \mathcal{P}^{t}_{CGD} \cup \mathcal{P}^{t}_{CGU} \cup \mathcal{P}^{t}_{CGR} \cup \mathcal{P}^{t}_{CGA}$\;
			$t \pluseq 1$\;
		}
		\Comment*[f]{IPP-solutioning:}\;
		Formulate the IPP using $\mathcal{P}^T_{LP}$ and $\mathcal{F}$\;
		$\mathcal{P}^T_{IP} \gets$ Solve the IPP using a branch-and-cut algorithm based MIP solver until its run-time becomes $\geq Th_{ipt}$\;
		\vspace{3pt}\Comment*[f]{Termination of the Optimization Engine:}\;
		\If{$Z^T_{IP} \left(\text{cost of }\mathcal{P}^T_{IP}\right) = Z^T_{LP} \left(\text{cost of }\mathcal{P}^T_{LP}\right)$}{
			$\mathcal{P}^\star_{IP} \gets \mathcal{P}^T_{IP}$\;
			\textbf{Break}\;
		}
		$T \pluseq 1$\;
	}
	\textbf{return} $\mathcal{P}_{IP}^{\star}$\;
	\caption{Procedure for the Optimization Engine}
	\label{algo:optEng}
\end{algorithm}
Its input involves the given flight set $\mathcal{F}$; the generated IFS $\mathcal{P}_{IFS}$; the pre-defined termination parameters-- $Th_{cost}$ \& $Th_t$ (for CG-driven LPP-solutioning) and $Th_{ipt}$ (for IPP-solutioning); and the sub-routines for Legal Crew Pairing Generator ($\texttt{Pairing\_Gen()}$) and the four CG strategies ($\texttt{CGD()},~\texttt{CGU()},~\texttt{CGR()}~$and$~\texttt{CGA()}$) in the proposed CG heuristic. In each LPP-IPP interaction of the Optimization Engine, first, the CG-driven LPP-solutioning is executed (lines 3-25). It entails several LPP iterations (tracked by $t$), in each of which the first step is to formulate the \textit{primal} using $\mathcal{F}$ and the respective input pairing set. This input pairing set is:
\begin{itemize}
	\item $\mathcal{P}_{IFS}$, if the first LPP iteration ($t=1$) of the first LPP-IPP interaction ($T=1$) is being executed (lines 5-6).
	\item $\mathcal{P}^{T-1}_{IP}$, if the first LPP iteration ($t=1$) of any subsequent LPP-IPP interaction ($T>1$) is being executed (lines 7-8).
	\item $\mathcal{P}^{t-1}_{CG} \cup \mathcal{P}^{t-1}_{LP}$, if any subsequent LPP iteration ($t>1$) of any LPP-IPP interaction ($T \geq 1$) is being executed (lines 9-11).
\end{itemize}
Once the primal is formulated, it is solved using the corresponding LP solver to obtain the current optimal LPP solution, constituted by $\mathcal{P}^{t}_{LP}~$and$~X^t_{LP}$ (line 12). Subsequently, the termination criterion of CG-driven LPP-solutioning is checked (lines 13-16). If it is terminated, then the current LPP solution $\mathcal{P}^{t}_{LP}$ is fetched as the final LPP solution $\mathcal{P}^T_{LP}$ of this LPP-IPP interaction. If not, then a \textit{dual} is formulated using $\mathcal{P}^t_{LP}$ and $\mathcal{F}$ (line 17), which is then solved using the corresponding LP solver to obtain the current optimal dual vector $Y^t$ (line 18). Using the current $\mathcal{P}^{t}_{LP}$, $X^t_{LP}$ and $Y^t$, a fresh set of pairings $\mathcal{P}^t_{CG}$ is obtained using the CG heuristic, which is constituted by the new pairing sets from the four underlying CG strategies (lines 19-23). At the end of the LPP iteration $t$, the fresh set of pairings $\mathcal{P}^{t}_{CG}$ is combined with the current $\mathcal{P}^{t}_{LP}$ to serve as input pairing set for the subsequent LPP iteration ($t \pluseq 1$). Once this submodule is terminated, the resulting $\mathcal{P}^{T}_{LP}$ is passed over to the IPP-solutioning for its integerization, wherein, the MIP solver is used to obtain the IPP solution $\mathcal{P}^{T}_{IP}$ (lines 26 and 27). In that, the pre-defined $Th_{ipt}$ time-limit is used to terminate the MIP-search, if it does not self-terminate a priori. Subsequently, the resulting $\mathcal{P}^{T}_{IP}$ is passed back to the CG-driven LPP-solutioning for the next LPP-IPP interaction ($T \pluseq 1$), or returned as the final integer solution $\mathcal{P}^\star_{IP}$, depending upon the termination condition of the Optimization Engine (lines 28-32). In that, if the cost of $\mathcal{P}^T_{IP}~\left(Z_{IP}^T\right)$, matches the cost of $\mathcal{P}^T_{LP}~\left(Z_{LP}^{p,T}\right)$, then the Optimization Engine is terminated.

\section{Computational Experiments} \label{sec:exp}
This section first presents the test cases and the computational setup, used to investigate the utility of $AirCROP$, its modules, and their interactions. Subsequently, the settings of parameters involved in different modules of $AirCROP$ are presented. Lastly, the experimental results are discussed.

\subsection{Test Cases and Computational Setup} \label{sec:exp-testcases}
The real-world airline test cases, used for experimentation, are detailed in Table~\ref{tab:testcases}. Each of these test cases involves a weekly flight schedule, and have been provided by the research consortium's industrial sponsor (from the networks of US-based airlines).
\begin{table}[pos=htbp,align=\centering]
	\small 
	\caption{Real-world airline test cases used in this research work}
	\begin{center}
		\begin{tabular}{ccccc}
			\toprule
			\textbf{Test Cases} & \textbf{$\#$Flights} & \textbf{$\#$Crew Bases} & \textbf{$\#$Airports}& \textbf{$\#$Legal Duties} \\ \midrule
			TC-1 & 3202 & 15 & 88 & 454205 \\
			TC-2 & 3228 & 15 & 88 & 464092 \\
			TC-3 & 3229 & 15 & 88 & 506272 \\
			TC-4 & 3265 & 15 & 90 & 446937 \\
			TC-5 & 4212 & 15 & 88 & 737184 \\
			\bottomrule
		\end{tabular}
		\label{tab:testcases}
	\end{center}
\end{table}
\begin{figure}[pos=htbp,align=\justify]
	\centering
	\begin{subfigure}[t]{0.45\textwidth}
		\centering \includegraphics[width=\columnwidth, keepaspectratio]{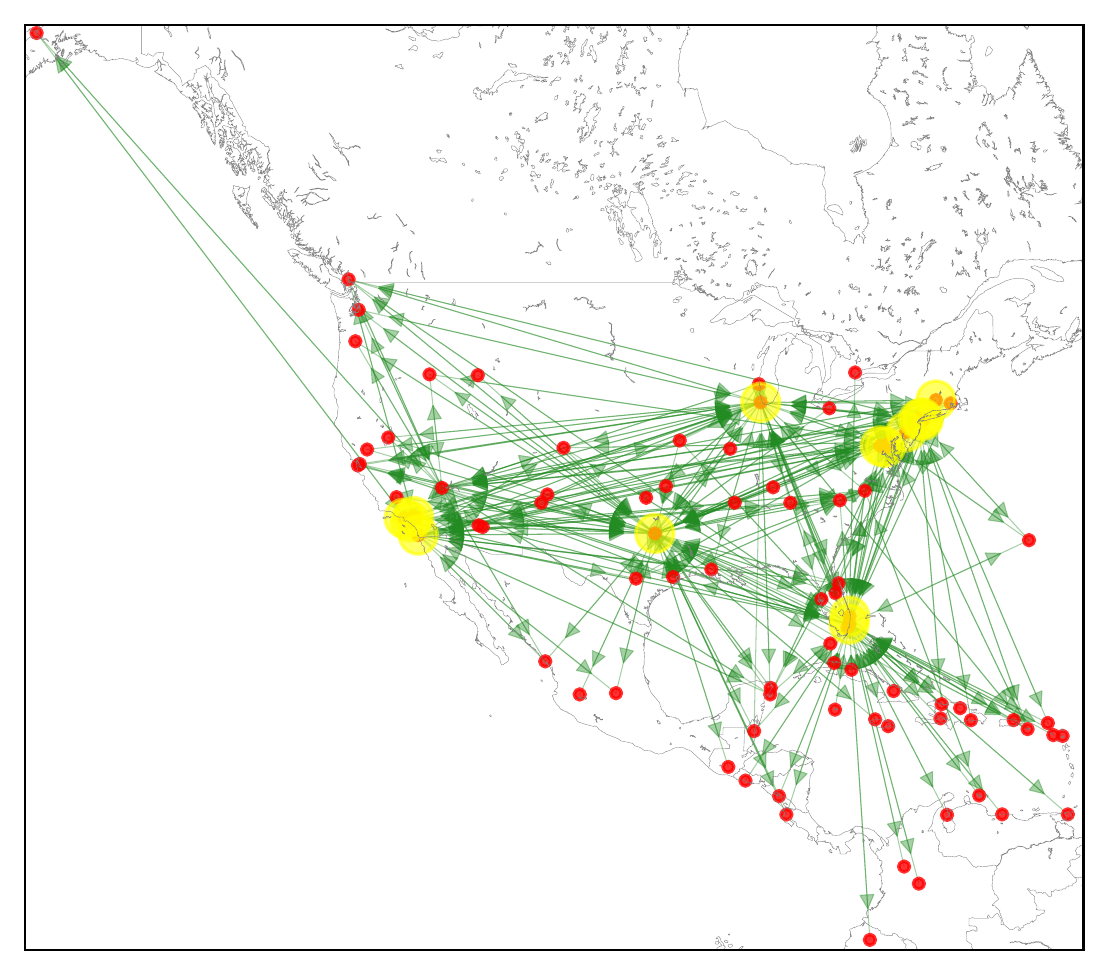}
		\caption{}
		\label{fig:flightmap}
	\end{subfigure}
	\begin{subfigure}[t]{0.415\textwidth}
		\centering \includegraphics[width=\columnwidth, keepaspectratio]{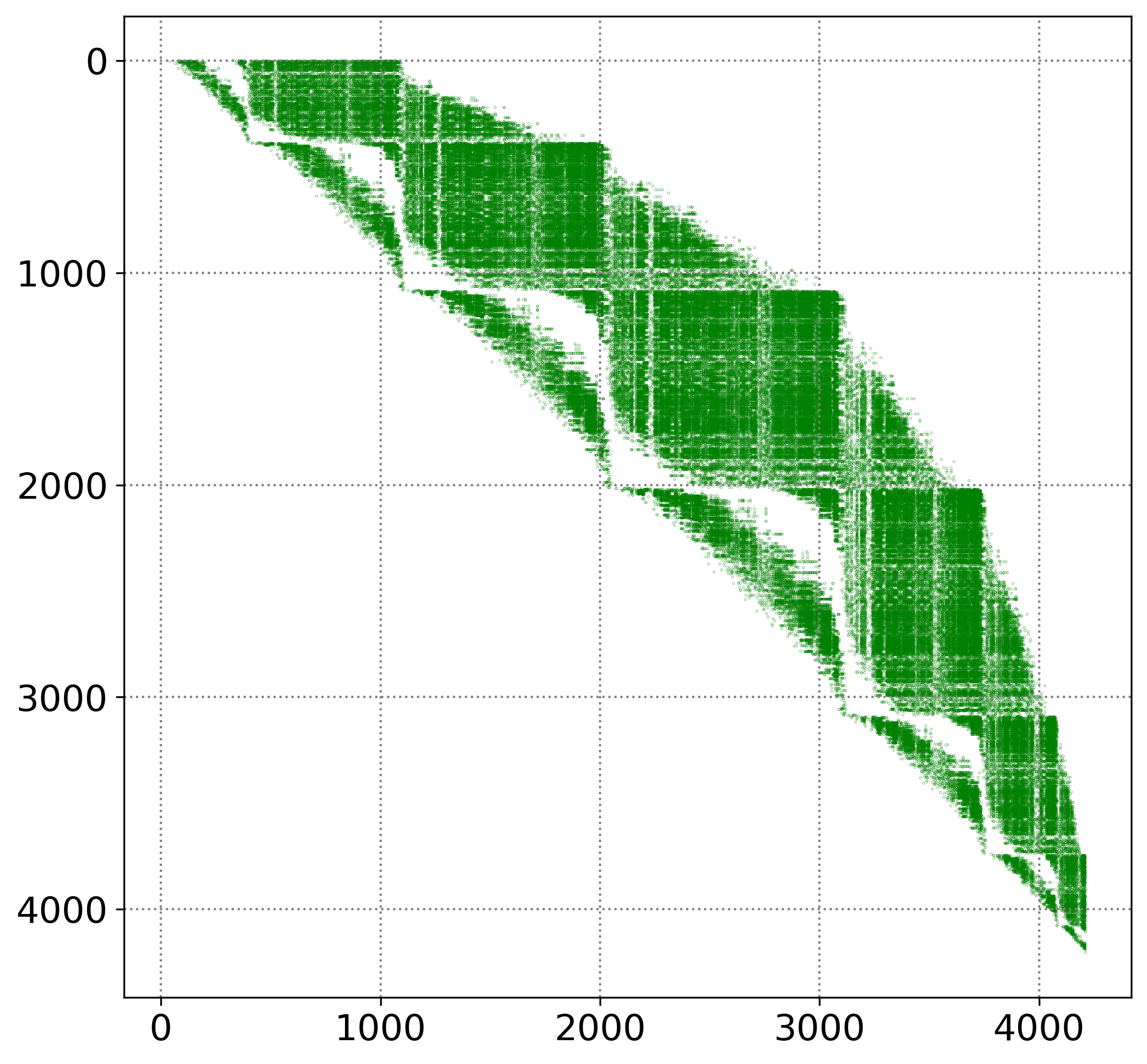}
		\caption{}
		\label{fig:plotTCs}
	\end{subfigure}
	\caption{(a) Geographical representation of TC-5 flight network, where the red nodes, green edges and yellow nodes represent the airports, scheduled flights and crew bases, respectively, and (b) \textit{legal} flight-connections, each represented by a point in the plot, where for a flight marked on the y-axis, the connecting flight is marked on the x-axis.}
	\label{fig:TCs}
\end{figure}
The columns in Table~\ref{tab:testcases}, in order of their occurrence, highlight the notations for the different test cases; the number of its constituent flights; the number of constituent crew bases; and the total number of legal duties involved, respectively. It is critical to recognize that the \textit{challenge associated with solutioning of these test cases, depends not just on the number of flights involved but also to the fact that these flights are part of complex flight networks, characterized by a multiplicity of hubs as opposed to a single hub, and multiplicity of crew bases as opposed to a single crew base}. In that, the number of legal pairings possible, grow exponentially with the number of hubs and crew bases. As a sample instance, the geographical representation of the flight network associated with TC-5, and the \textit{legal} flight connections involved in it, are portrayed in Figure~\ref{fig:TCs}. Notably, in Figure~\ref{fig:flightmap}, the presence of multiple hub-and-spoke subnetworks and multiple crew bases (highlighted in yellow color) is evident. Furthermore, the pattern visible in Figure~\ref{fig:plotTCs} could be attributed to the (minimum and maximum) limits on the sit-time and overnight-rest constraints. For instance, a flight, say $f_{500}$, has legal connections only with those flights that depart from the arrival airport of $f_{500}$, and whose departure-time gap (difference between its departure-time and the arrival time of $f_{500}$) lies within the minimum and maximum allowable limits, of the sit-time or the overnight-rest.
\par All the experiments in this research have been performed on an HP Z640 Workstation, which is powered by two Intel$^\circledR$ Xeon$^\circledR$ E5-2630v3 processors, each with 16 cores at 2.40 GHz, and 96 GBs of RAM. All codes related to the $AirCROP$ have been developed using the Python scripting language in alignment with the Industrial sponsor's larger vision and preference. Furthermore: 
\begin{itemize}
	\item the interior-point method from Gurobi Optimizer 8.1.1 \citep{gurobi} is used to solve the primal in the CG-		driven LPP-solutioning submodule.
	\item the interior-point method \citep{andersen2000mosek} from SciPy's linprog library \citep{scipy} is used to solve the dual in the CG-driven LPP-solutioning submodule.
	\item the branch-and-cut algorithm based MIP solver from Gurobi Optimizer 8.1.1 is used to solve the IPP in the Initial Feasible Solution Generator and the IPP-solutioning submodule.   
	\item an $AirCROP$-run, in principle, terminates when the cost of the IPP solution matches the cost of its input LPP solution in a particular LPP-IPP interaction. However, for practical considerations on the time-limit, an $AirCROP$-run is allowed to terminate if the IPP and LPP costs do not conform with each other even after 30 LPP-IPP interactions are over, or 30 hours of total run-time is elapsed.
\end{itemize}

\subsection{Parameter Settings} \label{sec:parameter-setting}
The settings of the parameters associated with different modules and submodules of the $AirCROP$ are, as highlighted below. 
\begin{itemize}
	\item \textit{Initial Feasible Solution Generator}: here, the proposed IPDCH involves the decomposition parameter $K$, which regulates the size of flight subsets formed in each of IPDCH-iteration. As mentioned before, the setting of $K$ is dependent on the characteristics of input flight dataset and the configuration of available computational resources. Here, the aim is to cover all given flights in a time-efficient manner. Hence, it is important to understand the effect of setting of $K$ on the time-performance of IPDCH, which is highlighted below.
	\begin{itemize}
		\item For a relatively lower value of $K$, smaller flight subsets with lesser number of legal flight-connections would be formed in each IPDCH-iteration, leading to coverage of relatively lesser number of unique flights in each of them. Though, this by itself is not a challenge, but this would necessitate a significant number of additional IPDCH-iterations (and the respective run-time), since the number of unique flights covered per IPDCH-iteration, which by construct reduces with the iterations, would get further reduced with relatively smaller flight subsets.
		\item On the flip side, for a relatively higher value of $K$, bigger flight subsets would be formed that would lead to coverage of higher number of unique flights per IPDCH-iteration. Though, this may reduce the total number of IPDCH-iterations required to generate the desired IFS, the overall run-time of the IPDCH may increase drastically. The rationale being that with bigger flight subset in each IPDCH-iteration, the number of possible legal pairings would increase drastically, leading to huge run-time for their generation as well as for the subsequent MIP-search.
	\end{itemize}
	The above considerations suggest that $K$ should be reasonably-sized. Considering the given computational resources and the results of initial exploration around the possible number of pairings for differently-sized flight sets, the value of $K$ in each IPDCH-iteration is guided by a random integer between one-eighth and one-fourth of the size of the input flight set $\mathcal{F}$. It may be noted that this setting of $K$ has been selected considering the scale and complexity of the given test cases, and it needs to be re-visited if the scale and complexity of the flight network changes drastically.
	\item \textit{CG-driven LPP-solutioning}: The parameters involved in the termination criterion for this submodule-- $Th_{cost}$ \& $Th_t$, are set as 100 USD \& 10 iterations respectively, to achieve an LPP solution with a sufficiently good cost in a reasonably good time. Moreover, the sensitivity of these parameters towards the $AirCROP$'s performance is discussed in Section~\ref{sec:computeTermSetting}. Moreover, the effect of the parameter-- size of $\mathcal{P}^t_{CG}$, on the performance of this submodule (the final LPP solution's cost and required run-time), and the demand on the computational resources (dominantly, RAM) is highlighted below. 
	\begin{itemize}
		\item for a relatively small-sized $\mathcal{P}^t_{CG}$, the alternative pairings available to foster further cost improvement shall be quite limited, amounting to smaller cost benefits in each phase of the CG-driven LPP-solutioning. This would necessitate far more LPP-IPP interactions, to reach the near-optimal cost. This pre se is not a challenge, however, significant amount of additional run-time may be required, since: (a) each call for CG-driven LPP-solutioning demands a minimum of 10 LPP iterations, before it could be terminated, (b) such calls when invoked repeatedly, may consume significant run-time, yet, without reasonable cost benefit. 
		\item On the other hand, for a very large-sized $\mathcal{P}_{CG}^{t}$, though the potential for significant cost benefits may exist, the demand on the RAM may become overwhelming for any CG-driven LPP-solutioning phase to proceed. 
	\end{itemize}
	The above considerations suggest that the size of $\mathcal{P}_{CG}^{t}$ may neither be too small nor too large. Factoring these, the experiments here aim at $\mathcal{P}_{CG}^{t}$ sized approximately of a million pairings (significant size, yet, not overwhelming for 96 GB RAM). Furthermore, for a search that is not biased in favor of any particular CG strategy, the number of pairings from each CG strategy towards the overall CG heuristic are kept equable.
	\item \textit{IPP-solutioning}: As mentioned before, the MIP-search on a large-scale IPP is time-intensive. Hence, the termination parameter-- $Th_{ipt}$, that restricts the run-time of any IPP-solutioning phase if not self-terminated a priori, is reasonably set as 20 minutes, and its sensitivity on the $AirCROP$'s performance is discussed in Section~\ref{sec:computeTermSetting}.
\end{itemize}

\subsection{Results \& Observations} \label{results}
This section presents the experimental results and associated inferences, in the order highlighted below.
\begin{enumerate}
	\item The performance of the proposed $AirCROP$ on the given test cases with the aforementioned parameter settings is discussed.
	\item The phenomenon referred to as \textit{performance variability} \citep{lodi2013performance} is discussed in the context of $AirCROP$. This aspect is pertinent  since some variability in performance (even for the same random seed) is inevitable owing to $AirCROP$'s reliance on the mathematical programming solvers, which over the different runs may pick different permutations of the rows (flight-coverage) or columns (pairings).
	%
	%
	\item The impact of the initialization methods: (a) the proposed IPDCH, (b) an Enhanced-DFS heuristic, earlier proposed by the authors~\citep{aggarwal2018large}, and (c) a commonly adopted \textit{Artificial Pairings} method \citep{hoffman1993solving, vance1997heuristic}, on the final performance of $AirCROP$ is investigated.
	%
	%
	%
	\item The sensitivity of $AirCROP$'s performance to the termination parameters in  the Optimization Engine's submodules (CG-driven LPP-solutioning and IPP-solutioning) has been discussed.
\end{enumerate}

\subsubsection{\textit{AirCROP}'s Performance} \label{sec:computeBaseConfig}
The results of the $AirCROP$-runs on the given test cases (TC-1 to TC-5) with the aforementioned parameter settings are reported in Table~\ref{tab:AirCROP}. In that, for each test case:
\begin{itemize}
	\item the first row marked by ``$\mathcal{P}_{IFS}$'' highlights the cost associated with the IFS that initializes the $AirCROP$-run and the run-time consumed in its generation.
	\item the subsequent rows present the results of the LPP-IPP interactions (marked by the counter $T$). In that, for a particular $T$, the cost of the LP-solution passed on for its integerization and the associated time are highlighted. Also the cost of the IP-solution returned and the associated time are highlighted. Here, the unit of cost is USD, and the time corresponds to the HH:MM format.
	\item the final crew pairing solution ($\mathcal{P}^\star_{IP}$) is highlighted in the last row (emboldened) marked by ``Final Solution''.
\end{itemize}
It may be noted that the experimental results in the subsequent sections are presented in the same format, unless any digression is specifically highlighted.
\begin{table}[pos=htbp,align=\centering]
	\scriptsize
	\caption{$AirCROP$'s performance$^\ast$ on the given test cases}
	\begin{tabular}{cccccccccccc}
		\toprule
		\multicolumn{2}{c}{\textbf{LPP-IPP}} & \multicolumn{2}{c}{\multirow{2}{*}{\textbf{TC-1}}} & \multicolumn{2}{c}{\multirow{2}{*}{\textbf{TC-2}}} & \multicolumn{2}{c}{\multirow{2}{*}{\textbf{TC-3}}} & \multicolumn{2}{c}{\multirow{2}{*}{\textbf{TC-4}}} & \multicolumn{2}{c}{\multirow{2}{*}{\textbf{TC-5}}} \\
		\multicolumn{2}{c}{\textbf{Interactions}} &  &  &  &  &  &  &  &  &  &  \\ \cmidrule(r){1-2} \cmidrule(lr){3-4} \cmidrule(lr){5-6} \cmidrule(lr){7-8} \cmidrule(lr){9-10} \cmidrule(l){11-12}
		$T$ & $\mathcal{P}_{LP}^T/\mathcal{P}_{IP}^T$ & \textbf{Cost} & \textbf{Time} & \textbf{Cost} & \textbf{Time} & \textbf{Cost} & \textbf{Time} & \textbf{Cost} & \textbf{Time} & \textbf{Cost} & \textbf{Time} \\ \midrule
		\multicolumn{2}{c}{$\mathcal{P}_{IFS}$} & 85893202 & 00:05 & 81950079 & 00:05 & 51552744 & 00:03 & 131716653 & 00:08 & 89690776 & 00:06 \\ \midrule
		\multirow{2}{*}{1} & $\mathcal{P}_{LP}^1$ & 3468349 & 03:56 & 3493986 & 03:56 & 3483057 & 05:18 & 3595565 & 03:27 & 4583484 & 07:48 \\ \cmidrule{2-12}
		& $\mathcal{P}_{IP}^1$ & 3689420 & 00:20 & 3715798 & 00:20 & 3697204 & 00:20 & 3807233 & 00:20 & 4930789 & 00:20 \\ \midrule
		\multirow{2}{*}{2} & $\mathcal{P}_{LP}^2$ & 3467837 & 02:18 & 3494675 & 01:19 & 3484645 & 02:42 & 3600195 & 01:17 & 4588740 & 02:49 \\ \cmidrule{2-12}
		& $\mathcal{P}_{IP}^2$ & 3557615 & 00:20 & 3587139 & 00:20 & 3590336 & 00:20 & 3679138 & 00:20 & 4734553 & 00:20 \\ \midrule
		\multirow{2}{*}{3} & $\mathcal{P}_{LP}^3$ & 3469591 & 00:47 & 3495254 & 01:22 & 3486614 & 01:59 & 3600813 & 01:16 & 4592143 & 01:46 \\ \cmidrule{2-12}
		& $\mathcal{P}_{IP}^3$ & 3518161 & 00:02 & 3546777 & 00:02 & 3523538 & 00:02 & 3639313 & 00:01 & 4654258 & 00:20 \\ \midrule
		\multirow{2}{*}{4} & $\mathcal{P}_{LP}^4$ & 3471619 & 01:13 & 3496797 & 00:57 & 3491000 & 01:13 & 3601168 & 01:27 & 4593422 & 02:17 \\ \cmidrule{2-12}
		& $\mathcal{P}_{IP}^4$ & 3489534 & 00:01 & 3505941 & 00:01 & 3496142 & 00:01 & 3621723 & 00:01 & 4634187 & 00:01 \\ \midrule
		\multirow{2}{*}{5} & $\mathcal{P}_{LP}^5$ & 3472403 & 00:31 & 3497106 & 00:23 & 3490420 & 00:56 & 3604082 & 00:37 & 4594282 & 02:14 \\ \cmidrule{2-12}
		& $\mathcal{P}_{IP}^5$ & 3484783 & 00:01 & 3497106 & 00:01 & 3490420 & 00:01 & 3612845 & 00:01 & 4617838 & 00:01 \\ \midrule
		\multirow{2}{*}{6} & $\mathcal{P}_{LP}^6$ & 3473238 & 00:30 &  &  &  &  & 3604753 & 00:28 & 4595481 & 01:53\\ \cmidrule{2-12}
		& $\mathcal{P}_{IP}^6$ & 3473238 & 00:01 &  &  &  &  & 3604753 & 00:01 & 4615272 & 00:01 \\ \midrule
		\multirow{2}{*}{7} & $\mathcal{P}_{LP}^7$ &  &  &  &  &  &  &  &  & 4596466 & 01:12 \\ \cmidrule{2-12}
		& $\mathcal{P}_{IP}^7$ &  &  &  &  &  &  &  &  & 4600428 & 00:01 \\ \midrule
		\multirow{2}{*}{8} & $\mathcal{P}_{LP}^8$ &  &  &  &  &  &  &  &  & 4595613 & 01:42 \\ \cmidrule{2-12}
		& $\mathcal{P}_{IP}^8$ &  &  &  &  &  &  &  &  & 4595613 & 00:01 \\ \midrule
		\multicolumn{2}{c}{\textbf{Final Solution}} & \textbf{3473238} & \textbf{10:05} & \textbf{3497106} & \textbf{08:46} & \textbf{3490420} & \textbf{12:55} & \textbf{3604753} & \textbf{09:24} & \textbf{4595613} & \textbf{22:52} \\
		\bottomrule
	\end{tabular}
	\label{tab:AirCROP}
	\\ \vspace{2pt} \footnotesize{$^\ast$All values in the ``Cost'' columns are in USD, and all corresponding real values are rounded-off to the next integer values. All values in the ``Time'' columns are in HH:MM format, and all corresponding seconds' values are rounded-off to the next minute values.}
\end{table}
\par The above results have been tested by the research consortium's industrial sponsor, and verified to be highly-competitive compared to the best practice solutions known, for different test cases. In general, the obtained solutions have been found to be superior by about 1.5 to 3.0\% in terms of the \textit{hard cost}, which reportedly is one of the most important solution quality indicator. For reference, a comparison of the obtained solution vis-$\grave{a}$-vis the best known solution has been drawn for TC-5, in Table~\ref{tab:TC5-SolChars}, where a significant difference in terms of the size of pairings can be observed. Notably, the key features contributing to lower hard cost relate to presence of pairings with relatively lower - TAFB, overnight rests and meal cost. However, the obtained solution also entails more crew changes, some of which (involving aircraft change) negatively impact the soft cost. Hence, there appears to be a trade-off between the hard cost and the soft cost.
\begin{table}[pos=htbp,align=\centering]
	\footnotesize
	\caption{Salient features of $\mathcal{P}^\star_{IP}$ for TC-5: $AirCROP$'s solution vis-$\grave{a}$-vis the best practice solution}
	\begin{tabular}{lrr}
		\toprule
		\textbf{Features} & $\bm{AirCROP}$\textbf{'s solution} & \textbf{Best practice solution} \\
		\midrule
		$\#$ pairings & 926 & 783 \\
		$\#$ unique flights covered & 4,212 & 4,212 \\
		$\#$ deadhead flights & 3 & 3 \\
		$\#$ overnight-rests & 1,203 & 1,279 \\
		$\#$ crew changes & 1,002 & 825 \\
		$\#$ average crew changes per pairing & 1.082 & 1.054 \\
		Total TAFB (HH:MM) & 37444:54 & 38189:39 \\
		\cmidrule(lr){1-3}
		$\#$ pairings covering 2 flights & 303 & 205 \\
		$\#$ pairings covering 3 flights & 17 & 31 \\
		$\#$ pairings covering 4 flights & 170 & 95 \\
		$\#$ pairings covering 5 flights & 63 & 37 \\
		$\#$ pairings covering 6 flights & 202 & 153 \\
		$\#$ pairings covering 7 flights & 59 & 62 \\
		$\#$ pairings covering 8 flights & 83 & 90 \\
		$\#$ pairings covering 9 flights & 19 & 49 \\
		$\#$ pairings covering 10 flights & 8 & 45 \\
		$\#$ pairings covering 11 flights & 1 & 10 \\
		$\#$ pairings covering 12 flights & 1 & 5 \\
		$\#$ pairings covering 13 flights & 0 & 0 \\
		$\#$ pairings covering 14 flights & 0 & 1 \\
		\cmidrule(lr){1-3}
		Hotel cost (USD) & 166,240 & 176,170 \\
		Meal cost (USD) & 157,269 & 160,397 \\
		Hard cost (USD) & 340,671 & 350,818 \\
		Soft cost (USD) & 51,600 & 42,750 \\
		Actual flying cost (USD) & 4,203,342 & 4,203,342 \\
		Total cost (USD) & 4,595,613 & 4,596,910 \\
		\bottomrule
	\end{tabular}
	\label{tab:TC5-SolChars}
\end{table}

\subsubsection{Performance Variability in \textit{AirCROP}} \label{sec:computePerfVar}
These section investigates the sensitivity of $AirCROP$ with respect to the sources of variability over multiple runs, even for the same problem. This study assumes importance, considering that performance variability is rather inevitable when the mathematical programming based solution approaches are employed \citep{koch2011miplib}. As cited by \cite{lodi2013performance}, variability in the performance of LP \& MIP solvers may be observed on -- changing the computing platform (which may change the floating-point arithmetic), permuting the constraints/variables of the respective mathematical models, or changing the pseudo-random numbers' seed. These changes/permutations may lead to an entirely different outcome of the respective search algorithms (LP \& MIP), as highlighted below.
\begin{itemize}
	\item The root source for the performance variability in MIP is the \textit{imperfect tie-breaking}. A majority of the decisions to be taken during an MIP-search are dependent on-- the ordering of the candidates according to an interim \textit{score} as well as the selection of the \textit{best} candidate (one with the best score value). A \textit{perfect} score that could fully-distinguish between the candidates is not-known mostly due to the lack of theoretical knowledge, and even if it is known, it may be too expensive to compute\footnote{For instance, in a \textit{strong branching} scheme, the best variable to branch at each node is decided after simulating one-level of branching for each fractional variable, however, it is performed heuristically to make it a computationally-affordable task for MIP solvers \citep{linderoth2010milp}}. Furthermore, additional ties or tiebreaks could be induced by changing the floating-point operations, which inherently may change when the computing platform is changed. Amidst such an imperfect tie-breaking, the permutation of the variables/constraints changes the path within the MIP-search tree, leading to a completely different evolution of the algorithm with rather severe consequences.
	\item Depending upon the floating-point arithmetic or the sequence of variables loaded in an LPP, the performance of the simplex and interior-point methods may vary.
	\item The performance of the LP and MIP solvers is also affected by the choice of pseudo-random numbers' seed, wherever the decisions are made heuristically. For instance, an interior-point method in the LP solvers performs a (random) crossover to one of the vertices of the optimal face when the search reaches its (unique) center.
\end{itemize}
\begin{table}[pos=htbp,align=\centering]
	\scriptsize
	\caption{Performance variability assessment for $AirCROP$ on two test instances$^\ast$ (TC-2 and TC-5)}
	\resizebox{0.97\columnwidth}{!}{
		\begin{tabular}{ccccccccccccc}
			\toprule
			\multirow{2.5}{*}{\textbf{Test}} & \multicolumn{2}{c}{\textbf{LPP-IPP}} & \multicolumn{4}{c}{\textbf{Runs with performance variability}} & \multicolumn{6}{c}{\textbf{Runs without performance variability}} \\ \cmidrule(lr){4-7}	\cmidrule(l){8-13}
			\multirow{2.5}{*}{\textbf{Case}} & \multicolumn{2}{c}{\textbf{Interactions}} & \multicolumn{2}{c}{\textbf{Run-1}} & \multicolumn{2}{c}{\textbf{Run-2}} & \multicolumn{2}{c}{\textbf{Run (Seed-$\alpha$)}} & \multicolumn{2}{c}{\textbf{Run (Seed-$\beta$)}} & \multicolumn{2}{c}{\textbf{Run (Seed-$\gamma$)}} \\ \cmidrule(r){2-3} \cmidrule(lr){4-5} \cmidrule(lr){6-7} \cmidrule(lr){8-9} \cmidrule(lr){10-11} \cmidrule(l){12-13}
			& $T$ & $\mathcal{P}_{LP}^T/\mathcal{P}_{IP}^T$ & \textbf{Cost} & \textbf{Time} & \textbf{Cost} & \textbf{Time} & \textbf{Cost} & \textbf{Time} & \textbf{Cost} & \textbf{Time} & \textbf{Cost} & \textbf{Time} \\ \midrule
			\multirow{22.5}{*}{\textbf{TC-2}} & \multicolumn{2}{c}{$\mathcal{P}_{IFS}$} & 81950079 & 00:05 & 74533686 & 00:04 & 129221508 & 00:08 & 114054265 & 00:07 & 52515476 & 00:04 \\ \cmidrule{2-13}
			& \multirow{2}{*}{1} & $\mathcal{P}_{LP}^1$ & 3493986 & 03:56 & 3494580 & 03:57 & 3495054 & 03:51 & 3493757 & 03:52 & 3493909 & 03:52 \\ \cmidrule{3-13}
			& & $\mathcal{P}_{IP}^1$ & 3715798 & 00:20 & 3746847 & 00:20 & 3769811 & 00:20 & 3711267 & 00:20 & 3722248 & 00:20 \\ \cmidrule{2-13}
			& \multirow{2}{*}{2} & $\mathcal{P}_{LP}^2$ & 3494675 & 01:19 & 3494540 & 02:18 & 3495311 & 01:57 & 3496733 & 01:57 & 3494657 & 03:02 \\ \cmidrule{3-13}
			& & $\mathcal{P}_{IP}^2$ & 3587139 & 00:20 & 3621066 & 00:20 & 3628514 & 00:20 & 3581978 & 00:20 & 3620745 & 00:20 \\ \cmidrule{2-13}
			& \multirow{2}{*}{3} & $\mathcal{P}_{LP}^3$ & 3495254 & 01:22 & 3496475 & 01:41 & 3497558 & 00:52 & 3499651 & 00:54 & 3496398 & 01:23 \\ \cmidrule{3-13}
			& & $\mathcal{P}_{IP}^3$ & 3546777 & 00:02 & 3555152 & 00:06 & 3566092 & 00:11 & 3536050 & 00:01 & 3551149 & 00:03 \\ \cmidrule{2-13}
			& \multirow{2}{*}{4} & $\mathcal{P}_{LP}^4$ & 3496797 & 00:57 & 3497750 & 01:36 & 3499237 & 01:26 & 3500818 & 01:03 & 3496069 & 01:37 \\ \cmidrule{3-13}
			& & $\mathcal{P}_{IP}^4$ & 3505941 & 00:01 & 3525600 & 00:01 & 3516807 & 00:01 & 3520552 & 00:01 & 3543236 & 00:02 \\ \cmidrule{2-13}
			& \multirow{2}{*}{5} & $\mathcal{P}_{LP}^5$ & 3497106 & 00:23 & 3498588 & 01:40 & 3500169 & 00:42 & 3500504 & 01:02 & 3496706 & 01:01 \\ \cmidrule{3-13}
			& & $\mathcal{P}_{IP}^5$ & 3497106 & 00:01 & 3498588 & 00:01 & 3517585 & 00:01 & 3500504 & 00:01 & 3501210 & 00:01 \\ \cmidrule{2-13}
			& \multirow{2}{*}{6} & $\mathcal{P}_{LP}^6$ &  &  &  &  & 3501523 & 00:43 &  &  & 3499063 & 00:41 \\ \cmidrule{3-13}
			& & $\mathcal{P}_{IP}^6$ &  &  &  &  & 3504085 & 00:01 &  &  & 3499063 & 00:01 \\ \cmidrule{2-13}
			& \multirow{2}{*}{7} & $\mathcal{P}_{LP}^7$ &  &  &  &  & 3502118 & 00:31 &  &  &  &  \\ \cmidrule{3-13}
			& & $\mathcal{P}_{IP}^7$ &  &  &  &  & 3502118 & 00:01 &  &  &  &  \\ \cmidrule{2-13}
			& \multicolumn{2}{c}{\textbf{Final Solution}} & \textbf{3497106} & \textbf{08:46} & \textbf{3498588} & \textbf{12:05} & \textbf{3502118} & \textbf{11:05} & \textbf{3500504} & \textbf{09:38} & \textbf{3499063} & \textbf{12:27} \\ \midrule \midrule
			\multirow{22.5}{*}{\textbf{TC-5}} & \multicolumn{2}{c}{$\mathcal{P}_{IFS}$} & 89690776 & 00:06 & 92080420 & 00:05 & 131443284 & 00:09 & 847887053 & 00:56 & 470430395 & 00:29 \\ \cmidrule{2-13}
			& \multirow{2}{*}{1} & $\mathcal{P}_{LP}^1$ & 4583484 & 07:48 & 4583476 & 08:00 & 4584525 & 07:28 & 4581988 & 08:47 & 4580130 & 07:36 \\ \cmidrule{3-13}
			& & $\mathcal{P}_{IP}^1$ & 4930789 & 00:20 & 4973580 & 00:20 & 4974341 & 00:20 & 4925863 & 00:20 & 4949616 & 00:20 \\ \cmidrule{2-13}
			& \multirow{2}{*}{2} & $\mathcal{P}_{LP}^2$ & 4588740 & 02:49 & 4588938 & 05:59 & 4589091 & 02:25 & 4584956 & 04:51 & 4584273 & 03:22 \\ \cmidrule{3-13}
			& & $\mathcal{P}_{IP}^2$ & 4734553 & 00:20 & 4765453 & 00:20 & 4782657 & 00:20 & 4749664 & 00:20 & 4753133 & 00:20 \\ \cmidrule{2-13}
			& \multirow{2}{*}{3} & $\mathcal{P}_{LP}^3$ & 4592143 & 01:46 & 4591571 & 02:35 & 4589952 & 02:14 & 4587812 & 03:02 & 4585046 & 03:40 \\ \cmidrule{3-13}
			& & $\mathcal{P}_{IP}^3$ & 4654258 & 00:20 & 4661078 & 00:20 & 4736313 & 00:20 & 4653279 & 00:20 & 4666390 & 00:20 \\ \cmidrule{2-13}
			& \multirow{2}{*}{4} & $\mathcal{P}_{LP}^4$ & 4593422 & 02:17 & 4595741 & 01:49 & 4591145 & 02:36 & 4589247 & 02:00 & 4588952 & 02:56 \\ \cmidrule{3-13}
			& & $\mathcal{P}_{IP}^4$ & 4634187 & 00:01 & 4624039 & 00:01 & 4654627 & 00:20 & 4614651 & 00:01 & 4628239 & 00:01 \\ \cmidrule{2-13}
			& \multirow{2}{*}{5} & $\mathcal{P}_{LP}^5$ & 4594282 & 02:14 & 4599006 & 01:14 & 4592463 & 02:03 & 4590573 & 01:05 & 4589577 & 02:02 \\ \cmidrule{3-13}
			& & $\mathcal{P}_{IP}^5$ & 4617838 & 00:01 & 4613385 & 00:01 & 4632708 & 00:02 & 4603938 & 00:01 & 4618710 & 00:01 \\ \cmidrule{2-13}
			& \multirow{2}{*}{6} & $\mathcal{P}_{LP}^6$ & 4595481 & 01:53 & 4598727 & 01:11 & 4593094 & 02:00 & 4591176 & 01:15 & 4589874 & 01:48 \\ \cmidrule{3-13}
			& & $\mathcal{P}_{IP}^6$ & 4615272 & 00:01 & 4605126 & 00:01 & 4625993 & 00:01 & 4591176 & 00:01 & 4607590 & 00:01 \\ \cmidrule{2-13}
			& \multirow{2}{*}{7} & $\mathcal{P}_{LP}^7$ & 4596466 & 01:12 & 4598412 & 01:39 & 4593431 & 01:04 &  &  & 4590674 & 01:24 \\ \cmidrule{3-13}
			& & $\mathcal{P}_{IP}^7$ & 4600428 & 00:01 & 4598412 & 00:01 & 4619643 & 00:01 &  &  & 4605058 & 00:01 \\ \cmidrule{2-13}
			& \multirow{2}{*}{8} & $\mathcal{P}_{LP}^8$ & 4595613 & 01:42 &  &  & 4594146 & 01:03 &  &  & 4591065 & 02:10 \\ \cmidrule{3-13}
			& & $\mathcal{P}_{IP}^8$ & 4595613 & 00:01 &  &  & 4594146 & 00:01 &  &  & 4591065 & 00:01 \\ \cmidrule{2-13}
			& \multicolumn{2}{c}{\textbf{Final Solution}} & \textbf{4595613} & \textbf{22:52} & \textbf{4598412} & \textbf{23:37} & \textbf{4594146} & \textbf{22:27} & \textbf{4591176} & \textbf{22:59} & \textbf{4591065} & \textbf{26:32} \\ \bottomrule
	\end{tabular}}
	\label{tab:PerVar}
	\\ \vspace{2pt} \footnotesize{$^\ast$All values in the ``Cost'' columns are in USD, and all the corresponding real values are rounded-off to the next integer values. All values in the ``Time'' columns are in HH:MM, and all the corresponding seconds' values are rounded-off to the next minute values.}
\end{table}
\par In the above background, the plausible reasons for variability in $AirCROP$'s performance are elaborated below.
\begin{itemize}
	\item \textit{Generation of new legal pairings using a parallel architecture: } in any LPP iteration $t$, new legal pairings are generated in parallel, by allocating the sub-processes to the idle-cores of the CPU. These sub-processes return their respective pairing sets as soon as they are terminated. This by itself is not a challenge, however, when the $AirCROP$ is re-run, the order in which these sub-processes terminate may not be same as before (as it depends on the state of the CPU), permuting the pairings in the cumulative pairing set $\mathcal{P}_{CG}^t$. This permuted pairing set, when fed as part of the input to the LP solver in the next LPP iteration, may lead to a different LPP solution, leading to a different outcome of the subsequent $AirCROP$'s search. To curb this, the pairings in the set that trigger the LP solver are sorted in \textit{lexicographical} order of their representative \textit{strings}. These strings are constructed from the indices of the flights covered in the corresponding pairings. For instance, the string corresponding to a pairing that covers flights $f_1$, $f_{10}$, $f_{100}$ \& $f_{200}$ is $1\_10\_100\_200$. Given that the pairings are distinct, the resulting strings are distinct too, allowing for a crisp sorting criterion and ensuring a fixed pairing sequence in each $AirCROP$-run.
	\item \textit{Numerical seed for generation of pseudo-random numbers}: variability may also be introduced if the numerical seed employed to generate pseudo-random numbers for use in the proposed modules or the utilized LP \& MIP solvers, varies. For instance, use of the default seed method of Python (i.e., the current time of the computing system) across different $AirCROP$ runs may lead to different pseudo-random numbers, each time. This in turn would trigger variability in the IFS generated by IPDCH (since the random selection of flights in each of its iterations, is impacted), and  the pairing set resulting from the CG heuristic (since each of the underlying CG strategy is impacted). Such variability could be negated by use of a fixed numerical seed, instead of a time dependent one. 
\end{itemize}
\par The intriguing questions for researchers could relate to the impact that presence or absence of causes of variability may have on the quality of $AirCROP$'s solutions, in terms of both cost and run-time. Table~\ref{tab:PerVar} attempts to shed light on these questions through empirical evidence for two test cases involving 3228 flights (TC-2) and 4212 flights (TC-5), respectively. In each of these test cases, the effect of variability is revealed through:
\begin{itemize}
	\item  two independent runs (Run-1 and Run-2), in each of which the causes of variability exist, that is: (a) the permutations of pairings generated using the parallel architecture is possible, and (b) the default seed method of Python, based on the time of the computing system applies.
	\item three independent runs, in each of which the causes of variability have been eliminated, that is: (a) the lexicographical order of the pairings is imposed, and (b) a fixed numerical seed has been fed for random number generation. For these runs, the numerical seeds are given by $\alpha=0$, $\beta=1$, and $\gamma=2$, respectively.
\end{itemize}
The key observations and inferences that could be drawn from each test case in Table~\ref{tab:PerVar} are highlighted below.
\begin{itemize}
	\item understandably, the Run-1 and Run-2 (corresponding to the same numerical seed), yield different cost solutions over different run-time. Importantly, the variation in cost (despite the presence of causes of variability) is not alarming, though significantly different run-times may be required.  
	\item each run (corresponding to Seed-$\alpha$, Seed-$\beta$, and Seed-$\gamma$, respectively) where the causes of variability have been negated, if repeated, yield the same cost solution in the same run-time though it has not been shown in the table for paucity of space. 
	\item the runs corresponding to the numerical seeds given by $\alpha$, $\beta$, and $\gamma$, respectively, differ solely due to the difference in the corresponding random numbers generated, and subsequently utilized. It can be observed that the change in numerical seed does not significantly affect the cost-quality of the final $AirCROP$ solution though the associated run-time may vary significantly. 
\end{itemize}
The fact that $AirCROP$ can offer final solutions with comparable cost quality, regardless of the presence or absence of causes of variability, endorses the robustness of the constitutive modules of the $AirCROP$. Also, the variation in run-time could be attributed to different \textit{search trajectories} corresponding to different permutations of variables or different random numbers. It may be noted that for the subsequent runs the lexicographical order of the pairings and a fixed numerical seed (Seed-$\alpha=0$) have been utilized. 

\subsubsection{Impact of Initialization on \textit{AirCROP}'s Performance} \label{sec:computeInit}
This section investigates the sensitivity of $AirCROP$ with respect to the cost quality of the initial solution and the run-time spent to obtain it. Towards it, the initial solution is obtained using three different methods (offering three input alternatives with varying cost and run-time) and the cost quality of $AirCROP's$ final solution alongside the necessary run-time is noted. 
\par Notably, in an initial attempt to generate IFS for large-scale CPOPs, the authors proposed a DFS algorithm based heuristic, namely, \textit{Enhanced-DFS} heuristic \citep{aggarwal2018large}. Its performance across the five test cases has been highlighted in Table~\ref{tab:IPDCHandEDFS}. In that, TC-1 emerges as an outlier owing to alarmingly high run-time, when compared to all other test cases.
\begin{table}[pos=htbp,align=\centering]
	\scriptsize
	\caption{Performance of Enhanced-DFS heuristic \citep{aggarwal2018large} for IFS generation. Here, the real valued ``Cost'' is rounded-off to the next integer value, and the seconds' in the ``Time'' column are rounded-off to the next minute values.}
	\begin{tabular}{cccc}
		\toprule
		\textbf{Test Cases}  & \textbf{Time (HH:MM)} & \textbf{Cost (USD)} & \textbf{\# Pairings} \\
		\midrule
		\textbf{TC-1}  & 01:48 & 3863669070 & 477617 \\ \midrule
		\textbf{TC-2}  & 00:02 & 167405376 & 26678 \\ \midrule
		\textbf{TC-3}  & 00:03 & 167967482 & 26871 \\ \midrule
		\textbf{TC-4}  & 00:13 & 1072078483 & 135269 \\ \midrule
		\textbf{TC-5}  & 00:04 & 325922318 & 51920 \\ \bottomrule
	\end{tabular}
	\label{tab:IPDCHandEDFS}
	\\ 
\end{table}
A plausible explanation behind this aberration is that TC-1 involves some flights with very few legal flight connections, and a DFS based algorithm may have to exhaustively explore several flight connections, to be able to generate an IFS with full flight coverage. The need to do away with reliance on DFS so as to have equable run-time across different data sets explains the motivation for: 
\begin{itemize} 
	\item proposition of IPDCH in this paper, which as highlighted in Section~\ref{sec:IFS}, relies on: (a) a divide-and-cover strategy to decompose the input flight schedule into sufficiently-small flight subsets, and (b) IP to find a lowest-cost pairing set that covers the maximum-possible flights for each of the decomposed flight subsets. 
	\item consideration of a commonly adopted \textit{Artificial Pairings} method \citep{vance1997heuristic}, that constructs a pairing set which covers all the flights, though some/all the pairings may not be legal. Hence, for this method the initial solution would be referred as $\mathcal{P}_{IS}$ instead of $\mathcal{P}_{IFS}$.
\end{itemize}
\begin{table}[pos=htbp,align=\centering]
	\scriptsize
	\caption{Performance assessment of $AirCROP$ on TC-1 and TC-5 when initialized using the proposed IPDCH, the Artificial Pairings method, and the Enhanced-DFS heuristic.}
	\resizebox{\columnwidth}{!}{%
		\begin{tabular}{cccccccccccccc}
			\toprule
			\multicolumn{2}{c}{\textbf{LPP-IPP}} & \multicolumn{6}{c}{\textbf{TC-1}} & \multicolumn{6}{c}{\textbf{TC-5}} \\
			\cmidrule(lr){3-8} \cmidrule(l){9-14}
			\multicolumn{2}{c}{\textbf{Interactions}} & \multicolumn{2}{c}{\textbf{Enhanced-DFS}} & \multicolumn{2}{c}{\textbf{IPDCH}} & \multicolumn{2}{c}{\textbf{Artificial Pairings}} & \multicolumn{2}{c}{\textbf{Enhanced-DFS}} &  \multicolumn{2}{c}{\textbf{IPDCH}} & \multicolumn{2}{c}{\textbf{Artificial Pairings}} \\ \cmidrule(r){1-2} \cmidrule(lr){3-4} \cmidrule(lr){5-6} \cmidrule(lr){7-8} \cmidrule(lr){9-10} \cmidrule(lr){11-12} \cmidrule(l){13-14}
			$T$ & $\mathcal{P}_{LP}^T/\mathcal{P}_{IP}^T$ & \textbf{Cost} & \textbf{Time} & \textbf{Cost} & \textbf{Time} & \textbf{Cost} & \textbf{Time} & \textbf{Cost} & \textbf{Time} & \textbf{Cost} & \textbf{Time} & \textbf{Cost} & Time \\ \midrule
			\multicolumn{2}{c}{$\mathcal{P}_{IFS}/\mathcal{P}_{IS}$} & 3863669070 & 01:48 & 74945982 & 00:05 & 14604919138 & $\approx$00:00 & 325922318 & 00:04 & 131443284 & 00:09 & 25409939785 & $\approx$00:00 \\ \midrule
			\multirow{2}{*}{1} & $\mathcal{P}_{LP}^1$ & 3463560 & 04:14 & 3465379 & 04:19 & 17589714 & 03:10 & 4583664 & 06:57 & 4584525 & 07:28 & 4585380 & 07:29 \\ \cmidrule(l){2-14}
			& $\mathcal{P}_{IP}^1$ & 3650828 & 00:20 & 3689312 & 00:20 & 17833718 & 00:20 & 4943531 & 00:20 & 4974341 & 00:20 & 4960813 & 00:20 \\ \midrule
			\multirow{2}{*}{2} & $\mathcal{P}_{LP}^2$ & 3464678 & 01:51 & 3466567 & 01:32 & 17589851 & 01:29 & 4586675 & 03:41 & 4589091 & 02:25 & 4589470 & 03:34 \\ \cmidrule(l){2-14}
			& $\mathcal{P}_{IP}^2$ & 17731125 & 00:20 & 3566415 & 00:20 & 3578030 & 00:20 & 4773342 & 00:20 & 4809348 & 00:20 & 4782657 & 00:20 \\ \midrule
			\multirow{2}{*}{3} & $\mathcal{P}_{LP}^3$ & 3466217 & 01:38 & 3467848 & 01:33 & 3466868 & 02:02 & 4586581 & 04:54 & 4589952 & 02:14 & 4593117 & 02:05 \\ \cmidrule(l){2-14}
			& $\mathcal{P}_{IP}^3$ & 3531694 & 00:13 & 3556499 & 00:20 & 3553432 & 00:20 & 4701607 & 00:20 & 4736313 & 00:20 & 4672696 & 00:20 \\ \midrule
			\multirow{2}{*}{4} & $\mathcal{P}_{LP}^4$ & 3467672 & 01:19 & 3468777 & 01:26 & 3467935 & 00:47 & 4589568 & 01:51 & 4591145 & 02:36 & 4593938 & 02:18 \\ \cmidrule(l){2-14}
			& $\mathcal{P}_{IP}^4$ & 3507987 & 00:01 & 3517901 & 00:01 & 3516376 & 00:02 & 4651824 & 00:20 & 4654627 & 00:20 & 4650449 & 00:06 \\ \midrule
			\multirow{2}{*}{5} & $\mathcal{P}_{LP}^5$ & 3469533 & 00:44 & 3468894 & 00:49 & 3468332 & 00:40 & 4591698 & 02:03 & 4592463 & 02:03 & 4596256 & 02:24 \\ \cmidrule(l){2-14}
			& $\mathcal{P}_{IP}^5$ & 3483690 & 00:01 & 3499531 & 00:01 & 3496156 & 00:01 & 4616605 & 00:01 & 4632708 & 00:02 & 4620903 & 00:01 \\ \midrule
			\multirow{2}{*}{6} & $\mathcal{P}_{LP}^6$ & 3469276 & 00:52 & 3469352 & 00:48 & 3469095 & 00:47 & 4591969 & 01:05 & 4593094 & 02:00 & 4597203 & 00:49 \\ \cmidrule(l){2-14}
			& $\mathcal{P}_{IP}^6$ & 3469276 & 00:01 & 3477354 & 00:01 & 3491947 & 00:01 & 4606253 & 00:01 & 4625993 & 00:01 & 4612164 & 00:01 \\ \midrule
			\multirow{2}{*}{7} & $\mathcal{P}_{LP}^7$ &  &  & 3469950 & 00:42 & 3469543 & 00:52 & 4592860 & 01:15 & 4593431 & 01:04 & 4597913 & 01:17 \\ \cmidrule(l){2-14}
			& $\mathcal{P}_{IP}^7$ &  &  & 3469950 & 00:01 & 3487562 & 00:01 & 4592860 & 00:01 & 4619643 & 00:01 & 4606368 & 00:01 \\ \midrule
			\multirow{2}{*}{8} & $\mathcal{P}_{LP}^8$ &  &  &  &  & 3470100 & 00:38 &  &  & 4594146 & 01:03 & 4597730 & 02:00 \\ \cmidrule(l){2-14}
			& $\mathcal{P}_{IP}^8$ &  &  &  &  & 3478057 & 00:01 &  &  & 4594146 & 00:01 & 4604551 & 00:01 \\ \midrule
			\multirow{2}{*}{9} & $\mathcal{P}_{LP}^9$ &  &  &  &  & 3470355 & 00:28 &  &  &  &  & 4597929 & 00:50 \\ \cmidrule(l){2-14}
			& $\mathcal{P}_{IP}^9$ &  &  &  &  & 3470355 & 00:01 &  &  &  &  & 4597929 & 00:01 \\ \midrule
			\multicolumn{2}{c}{\textbf{Final Solution}} & \textbf{3469276} & \textbf{13:22} & \textbf{3469950} & \textbf{12:18} & \textbf{3470355} & \textbf{12:00} & \textbf{4592860} & \textbf{23:13} & \textbf{4594146} & \textbf{22:27} & \textbf{4597929} & \textbf{23:57} \\ \bottomrule
	\end{tabular}}
	\label{tab:initialization}
	\\ \footnotesize{$^\ast$All values in the ``Cost'' columns are in USD, where the real values are rounded-off to the next integer values. All values in the ``Time'' columns are in HH:MM, where the seconds' values are rounded-off to the next minute values.}
\end{table}
A comparison of the above three methods has been drawn in Table~\ref{tab:initialization}, for TC-1 (posing challenge to Enhanced-DFS) and TC-5 (largest flight set). In that, besides the cost and run-time of the initial solution for each test case, the results of all the iterations of $AirCROP$ leading up to the final solution have been presented. The latter is done to shed light on whether $AirCROP's$ final solution cost quality strongly depends on the cost of the initial solution. The prominent observations from the Table~\ref{tab:initialization} include:
\begin{itemize}
	\item In terms of run-time: IPDCH could outperform the Enhanced-DFS, as its run-time happened to be less than ten minutes in both the test cases. The Artificial pairing method even out performs IPDCH, since its run-time happened to be in milliseconds (formatted to 0 minutes in the table). 
	\item In terms of initial cost: IPDCH could again outperform the Enhanced-DFS. This could be attributed to the use of IP to find a lowest-cost pairing set that covers the maximum-possible flights for each of the decomposed flight subsets. In contrast, the cost associated with the Artificial pairing method, is the worst. This is owing to a very high pseudo-cost attached to the pairings to offset their non-legality.  \end{itemize}
Critically, regardless of the significantly varying run-time and the initial cost associated with the three methods, the variation in the cost of the final solution offered by $AirCROP$ is not significant. This endorses the robustness of its constitutive modules.

\subsubsection{Impact of Termination Settings of Optimization Engine's Submodules on \textit{AirCROP}'s Performance} \label{sec:computeTermSetting}
This section investigates the sensitivity of $AirCROP$ to the termination parameter settings of the Optimization Engine's submodules, namely, LPP-solutioning and IPP-solutioning. The parameters involved in LPP-solutioning are $Th_{cost}$ and $Th_t$, while $Th_{ipt}$ is involved in IPP-solutioning. To assess their impact on $AirCROP's$ performance, experiments are performed with three different sets of parameter settings each, for both the submodules. \\\\
\noindent \textbf{Impact of Termination Settings of CG-driven LPP-solutioning:}\\
As mentioned earlier, the CG-driven LPP-solutioning is terminated if the cost-improvement per LPP iteration falls below the pre-specified threshold $Th_{cost}$ (in USD) over $Th_t$ number of successive LPP iterations. To achieve a reasonable balance between $AirCROP's$ run time on the one hand and the cost reduction of the crew pairing solution on the other hand, three different sets of parameter settings are chosen, and experimented with. These settings of $\{Th_{cost}, Th_t\}$ including $\{500,5\}$, $\{100,10\}$, and $\{50,15\}$ symbolize relaxed, moderate and strict settings, respectively, since the criterion for $AirCROP's$ termination gets more and more difficult as the settings change from $\{500,5\}$ to $\{50,15\}$. The results of the $AirCROP$-runs corresponding to these termination settings are reported in Table~\ref{tab:LPPTerm}, and the key observations are as highlighted below.
\begin{itemize}
	\item As the termination settings transition through relaxed, moderate and strict settings, the run-time to obtain the final solution increases, while the cost of the final solution decreases. An apparent exception to this trend is observed in TC-5 with the strict setting, but this could be explained by the fact that the upper limit of 30 hours set for $AirCROP's$ run time under practical considerations was exceeded during the fourth LPP-IPP interaction ($T=4$). It implies that due to the enforced termination in this particular case, $AirCROP$ could not fully utilize the potential for cost reduction.
	\item Despite the variation in the termination settings, the cost quality of $AirCROP's$ final solution does not vary as drastically, as its run time. For instance, as the settings switched from relaxed to moderate: an additional saving of 6384 USD could be achieved at the expense of additional 5:20 run time in the case of TC-2, while these indicators stand at 13388 USD and 10:25, respectively, in the case of TC-5. It can also be inferred that $\{Th_{cost}, Th_t\}$ set as $\{100,10\}$ possibly offers a fair balance between solution's cost quality and run time, and this explains why these settings have been used as the base settings for the experimental results presented in this paper, beginning with Table~\ref{tab:AirCROP} and ending with Table~\ref{tab:IPPTerm}.
\end{itemize}
It is important to recognize that as the termination settings for LPP-solutioning are made stricter, its run time is bound to increase. It is also fair to expect that the cost quality of the final solution may be better, though it cannot be guaranteed. Any such departures from the expected trend may be due to the dependence of the quality of the final solution on the quality of the IPP-solution for each $T$. In that, if an IPP-solution for a particular $T$ may largely fail to approach the lower bound set by the corresponding LPP-solution, it may negatively influence the cost quality obtained in subsequent LPP- and IPP-solutioning phases. While such a possibility remains, it did not surface in the experiments above. \\\\
\begin{table}[pos=htbp,align=\centering]
	\scriptsize
	\caption{Performance assessment of $AirCROP$ on TC-2 and TC-5, against three different termination settings (\textit{Relaxed}, \textit{Moderate} and \textit{Strict} Settings) of the CG-driven LPP-solutioning$^\ast$}
	\resizebox{\columnwidth}{!}{
		\begin{tabular}{cccccccccccccc}
			\toprule
			\multicolumn{2}{c}{\multirow{1.75}{*}{\textbf{LPP-IPP}}} & \multicolumn{6}{c}{\textbf{TC-2}} & \multicolumn{6}{c}{\textbf{TC-5 }} \\ \cmidrule(lr){3-8}	\cmidrule(l){9-14}
			\multicolumn{2}{c}{\multirow{1.75}{*}{\textbf{Interactions}}} & \multicolumn{2}{c}{\textbf{Relaxed Setting}} & \multicolumn{2}{c}{\textbf{Moderate Setting}} & \multicolumn{2}{c}{\textbf{Strict Setting}} & \multicolumn{2}{c}{\textbf{Relaxed Setting}} & \multicolumn{2}{c}{\textbf{Moderate Setting}} & \multicolumn{2}{c}{\textbf{Strict Setting}} \\
			& & \multicolumn{2}{c}{\textbf{$\bm{Th_{cost} =}$ 500, $\bm{Th_{t} =}$ 5}} & \multicolumn{2}{c}{\textbf{$\bm{Th_{cost} =}$ 100, $\bm{Th_{t} =}$ 10}} & \multicolumn{2}{c}{\textbf{$\bm{Th_{cost} =}$ 50, $\bm{Th_{t} =}$ 15}} & \multicolumn{2}{c}{\textbf{$\bm{Th_{cost} =}$ 500, $\bm{Th_{t} =}$ 5}} & \multicolumn{2}{c}{\textbf{$\bm{Th_{cost} =}$ 100, $\bm{Th_{t} =}$ 10}} & \multicolumn{2}{c}{\textbf{$\bm{Th_{cost} =}$ 50, $\bm{Th_{t} =}$ 15}} \\ \cmidrule(r){1-2} \cmidrule(lr){3-4} \cmidrule(lr){5-6} \cmidrule(lr){7-8} \cmidrule(lr){9-10} \cmidrule(lr){11-12}  \cmidrule(l){13-14}
			$T$ & $\mathcal{P}_{LP}^T/\mathcal{P}_{IP}^T$ & \textbf{Cost} & \textbf{Time} & \textbf{Cost} & \textbf{Time} & \textbf{Cost} & \textbf{Time} & \textbf{Cost} & \textbf{Time} & \textbf{Cost} & \textbf{Time} & \textbf{Cost} & \textbf{Time} \\ \midrule
			\multicolumn{2}{c}{$\mathcal{P}_{IFS}$} & 129221508 & 00:08 & 129221508 & 00:08 & 129221508 & 00:08 & 131443284 & 00:09 & 131443284 & 00:09 & 131443284 & 00:09 \\ \midrule
			\multirow{2}{*}{1} & $\mathcal{P}_{LP}^1$ & 3510231 & 01:12 & 3495054 & 03:51 & 3489337 & 08:21 & 4603984 & 02:43 & 4584525 & 07:28 & 4581711 & 10:05 \\ \cmidrule{2-14}
			& $\mathcal{P}_{IP}^1$ & 3844119 & 00:20 & 3769811 & 00:20 & 3698316 & 00:20 & 5165821 & 00:20 & 4974341 & 00:20 & 4946510 & 00:20 \\ \midrule
			\multirow{2}{*}{2} & $\mathcal{P}_{LP}^2$ & 3510105 & 00:26 & 3495311 & 01:57 & 3491725 & 03:36 & 4605049 & 01:12 & 4589091 & 02:25 & 4582977 & 09:32 \\ \cmidrule{2-14}
			& $\mathcal{P}_{IP}^2$ & 3729820 & 00:20 & 3628514 & 00:20 & 3607470 & 00:20 & 4962643 & 00:20 & 4782657 & 00:20 & 4780005 & 00:20 \\ \midrule
			\multirow{2}{*}{3} & $\mathcal{P}_{LP}^3$ & 3506864 & 00:35 & 3497558 & 00:52 & 3491685 & 04:15 & 4602283 & 01:09 & 4589952 & 02:14 & 4585457 & 05:56 \\ \cmidrule{2-14}
			& $\mathcal{P}_{IP}^3$ & 3659201 & 00:20 & 3566092 & 00:11 & 3578774 & 00:20 & 4818918 & 00:20 & 4736313 & 00:20 & 4678596 & 00:20 \\ \midrule
			\multirow{2}{*}{4} & $\mathcal{P}_{LP}^4$ & 3506644 & 00:32 & 3499237 & 01:26 & 3494201 & 02:29 & 4604535 & 00:52 & 4591145 & 02:36 & 4595692 & 03:34 \\ \cmidrule{2-14}
			& $\mathcal{P}_{IP}^4$ & 3606381 & 00:20 & 3516807 & 00:01 & 3540972 & 00:01 & 4727106 & 00:20 & 4654627 & 00:20 & 4624747 & 00:01 \\ \midrule
			\multirow{2}{*}{5} & $\mathcal{P}_{LP}^5$ & 3507647 & 00:29 & 3500169 & 00:42 & 3494409 & 02:36 & 4603253 & 00:47 & 4592463 & 02:03 &  &  \\ \cmidrule{2-14}
			& $\mathcal{P}_{IP}^5$ & 3559484 & 00:04 & 3517585 & 00:01 & 3527254 & 00:01 & 4683130 & 00:20 & 4632708 & 00:02 &  &  \\ \midrule
			\multirow{2}{*}{6} & $\mathcal{P}_{LP}^6$ & 3507101 & 00:20 & 3501523 & 00:43 & 3496498 & 01:00 & 4603093 & 00:45 & 4593094 & 02:00 &  &  \\ \cmidrule{2-14}
			& $\mathcal{P}_{IP}^6$ & 3547304 & 00:02 & 3504085 & 00:01 & 3496498 & 00:01 & 4681335 & 00:20 & 4625993 & 00:01 &  &  \\ \midrule
			\multirow{2}{*}{7} & $\mathcal{P}_{LP}^7$ & 3508166 & 00:18 & 3502118 & 00:31 &  &  & 4603638 & 00:46 & 4593431 & 01:04 &  &  \\ \cmidrule{2-14}
			& $\mathcal{P}_{IP}^7$ & 3517436 & 00:01 & 3502118 & 00:01 &  &  & 4651002 & 00:06 & 4619643 & 00:01 &  &  \\ \midrule
			\multirow{2}{*}{8} & $\mathcal{P}_{LP}^8$ & 3508502 & 00:17 &  &  &  &  & 4604073 & 00:44 & 4594146 & 01:03 &  &  \\ \cmidrule{2-14}
			& $\mathcal{P}_{IP}^8$ & 3508502 & 00:01 &  &  &  &  & 4634316 & 00:02 & 4594146 & 00:01 &  &  \\ \midrule
			\multirow{2}{*}{9} & $\mathcal{P}_{LP}^9$ &  &  &  &  &  &  & 4606250 & 00:28 &  &  &  &  \\ \cmidrule{2-14}
			& $\mathcal{P}_{IP}^9$ &  &  &  &  &  &  & 4614420 & 00:01 &  &  &  &  \\ \midrule
			\multirow{2}{*}{10} & $\mathcal{P}_{LP}^{10}$ &  &  &  &  &  &  & 4607534 & 00:17 &  &  &  &  \\ \cmidrule{2-14}
			& $\mathcal{P}_{IP}^{10}$ &  &  &  &  &  &  & 4607534 & 00:01 &  &  &  &  \\ \midrule
			\multicolumn{2}{c}{\textbf{Final Solution}} & \textbf{3508502} & \textbf{05:45} & \textbf{3502118} & \textbf{11:05} & \textbf{3496498} & \textbf{23:28} & \textbf{4607534} & \textbf{12:02} & \textbf{4594146} & \textbf{22:27} & \textbf{4624747} & \textbf{30:17} \\ \bottomrule
	\end{tabular}}
	\label{tab:LPPTerm}
	\\ \footnotesize{$^\ast$All values in the ``Cost'' columns are in USD, and all the corresponding real values are rounded-off to the next integer values. All values in the ``Time'' columns are in HH:MM, and all the corresponding seconds' values are rounded-off to the next minute values.}
\end{table}
%
\noindent \textbf{Impact of Termination Settings of IPP-solutioning:}\\
As mentioned before, integerization of an LPP solution using an MIP solver is extremely time-consuming, particularly for large-scale CPOPs, and more so those involving complex flight networks. Hence, from a practical perspective, the $AirCROP$ framework imposes a threshold on the upper time limit for IPP-solutioning (for any given $T$), namely $Th_{ipt}$, in case it does not self-terminate a priori. To investigate the impact of $Th_{ipt}$ on $AirCROP's$ performance, experiments are performed with three different settings, including, 00:20 (one-third of an hour), 00:40 (two-third of an hour), and 01:00 (an hour). The results are presented in Table~\ref{tab:IPPTerm}, and the key observations are as follows. In the case of TC-2, as the $Th_{ipt}$ is raised, the run-time to obtain the final solution increases, while the cost of the final solution decreases. However, there are exceptions to this trend in the case of TC-5. Notably, the cost quality of the final solution corresponding to $Th_{ipt}=$ 00:20 remains superior to that obtained for both $Th_{ipt}=$ 00:40 and 01:00. For these two settings, the quality of LPP-solution at $T=8$ turned worse compared to the case of $Th_{ipt}=$ 00:20, and the gap could not be bridged even in the subsequent LPP-IPP interaction ($T=9$). The worsening of LPP-solution could be attributed to the fact that LPP-solutioning relies on random number based heuristics, and the resulting pairing combinations may not necessarily offer lower cost within the pre-specified termination settings. 
\par Based on the above, it may be inferred that despite the changes in the termination parameter settings, $AirCROP$ is able to offer solutions with reasonably close cost quality, though significant variations in run time may be observed. It is also evident that even the lowest setting (desired from a practical perspective) for $Th_{ipt}=$ 00:20 offers a good balance between solution's cost quality and run time, and this explains why it has been used as the base setting for the experimental results presented in this paper.
\begin{table}[pos=htbp,align=\centering]
	\scriptsize
	\caption{Performance assessment of $AirCROP$ on TC-2 and TC-5, against three different termination settings ($Th_{ipt} =$ 00:20, 00:40 \& 01:00) of the IPP-solutioning$^\ast$}
	\resizebox{\columnwidth}{!}{
		\begin{tabular}{cccccccccccccc}
			\toprule
			\multicolumn{2}{c}{\multirow{1.5}{*}{\textbf{LPP-IPP}}} & \multicolumn{6}{c}{\textbf{TC-2 }} & \multicolumn{6}{c}{\textbf{TC-5 }} \\ \cmidrule(lr){3-8}	\cmidrule(l){9-14}
			\multicolumn{2}{c}{\multirow{0.75}{*}{\textbf{Interactions}}} & \multicolumn{2}{c}{\textbf{$\bm{Th_{ipt} =}$ 00:20}} & \multicolumn{2}{c}{\textbf{$\bm{Th_{ipt} =}$ 00:40}} & \multicolumn{2}{c}{\textbf{$\bm{Th_{ipt} =}$ 01:00}} & \multicolumn{2}{c}{\textbf{$\bm{Th_{ipt} =}$ 00:20}} & \multicolumn{2}{c}{\textbf{$\bm{Th_{ipt} =}$ 00:40}} & \multicolumn{2}{c}{\textbf{$\bm{Th_{ipt} =}$ 01:00}} \\ \cmidrule(r){1-2} \cmidrule(lr){3-4} \cmidrule(lr){5-6} \cmidrule(lr){7-8} \cmidrule(lr){9-10} \cmidrule(lr){11-12}  \cmidrule(l){13-14}
			$T$ & $\mathcal{P}_{LP}^T/\mathcal{P}_{IP}^T$ & \textbf{Cost} & \textbf{Time} & \textbf{Cost} & \textbf{Time} & \textbf{Cost} & \textbf{Time} & \textbf{Cost} & \textbf{Time} & \textbf{Cost} & \textbf{Time} & \textbf{Cost} & \textbf{Time} \\ \midrule
			\multicolumn{2}{c}{$\mathcal{P}_{IFS}$} & 129221508 & 00:08 & 129221508 & 00:08 & 129221508 & 00:08 & 131443284 & 00:09 & 131443284 & 00:09 & 131443284 & 00:09 \\ \midrule
			\multirow{2}{*}{1} & $\mathcal{P}_{LP}^1$ & 3495054 & 03:51 & 3495054 & 04:21 & 3495054 & 03:57 & 4584525 & 07:28 & 4584525 & 07:46 & 4584525 & 07:49 \\ \cmidrule{2-14}
			& $\mathcal{P}_{IP}^1$ & 3769811 & 00:20 & 3744301 & 00:40 & 3760028 & 01:00 & 4974341 & 00:20 & 4958532 & 00:40 & 4987497 & 01:00 \\ \midrule
			\multirow{2}{*}{2} & $\mathcal{P}_{LP}^2$ & 3495311 & 01:57 & 3497483 & 01:49 & 3495562 & 02:19 & 4589091 & 02:25 & 4585347 & 05:00 & 4588371 & 03:56 \\ \cmidrule{2-14}
			& $\mathcal{P}_{IP}^2$ & 3628514 & 00:20 & 3629401 & 00:40 & 3632875 & 01:00 & 4782657 & 00:20 & 4778465 & 00:40 & 4766924 & 01:00 \\ \midrule
			\multirow{2}{*}{3} & $\mathcal{P}_{LP}^3$ & 3497558 & 00:52 & 3497473 & 01:33 & 3494305 & 01:50 & 4589952 & 02:14 & 4589481 & 02:56 & 4588911 & 04:25 \\ \cmidrule{2-14}
			& $\mathcal{P}_{IP}^3$ & 3566092 & 00:11 & 3566247 & 00:40 & 3579899 & 01:00 & 4736313 & 00:20 & 4699845 & 00:40 & 4713402 & 01:00 \\ \midrule
			\multirow{2}{*}{4} & $\mathcal{P}_{LP}^4$ & 3499237 & 01:26 & 3500607 & 01:06 & 3495273 & 01:13 & 4591145 & 02:36 & 4590618 & 01:56 & 4591028 & 02:09 \\ \cmidrule{2-14}
			& $\mathcal{P}_{IP}^4$ & 3516807 & 00:01 & 3524672 & 00:01 & 3551863 & 00:05 & 4654627 & 00:20 & 4656611 & 00:40 & 4681015 & 01:00 \\ \midrule
			\multirow{2}{*}{5} & $\mathcal{P}_{LP}^5$ & 3500169 & 00:42 & 3501809 & 00:49 & 3496754 & 00:52 & 4592463 & 02:03 & 4591826 & 01:18 & 4591448 & 02:03 \\ \cmidrule{2-14}
			& $\mathcal{P}_{IP}^5$ & 3517585 & 00:01 & 3501809 & 00:01 & 3528564 & 00:01 & 4632708 & 00:02 & 4644467 & 00:15 & 4639287 & 00:29 \\ \midrule
			\multirow{2}{*}{6} & $\mathcal{P}_{LP}^6$ & 3501523 & 00:43 &  &  & 3496342 & 00:53 & 4593094 & 02:00 & 4592492 & 02:21 & 4591372 & 02:06 \\ \cmidrule{2-14}
			& $\mathcal{P}_{IP}^6$ & 3504085 & 00:01 &  &  & 3512692 & 00:01 & 4625993 & 00:01 & 4617694 & 00:01 & 4616944 & 00:01 \\ \midrule
			\multirow{2}{*}{7} & $\mathcal{P}_{LP}^7$ & 3502118 & 00:31 &  &  & 3497967 & 00:59 & 4593431 & 01:04 & 4594599 & 01:30 & 4594479 & 01:23 \\ \cmidrule{2-14}
			& $\mathcal{P}_{IP}^7$ & 3502118 & 00:01 &  &  & 3519996 & 00:01 & 4619643 & 00:01 & 4607261 & 00:01 & 4608085 & 00:01 \\ \midrule
			\multirow{2}{*}{8} & $\mathcal{P}_{LP}^8$ &  &  &  &  & 3498726 & 01:24 & 4594146 & 01:03 & 4595739 & 01:08 & 4595424 & 01:03 \\ \cmidrule{2-14}
			& $\mathcal{P}_{IP}^8$ &  &  &  &  & 3518299 & 00:01 & 4594146 & 00:01 & 4598624 & 00:01 & 4603634 & 00:01 \\ \midrule
			\multirow{2}{*}{9} & $\mathcal{P}_{LP}^9$ &  &  &  &  & 3499104 & 00:40 &  &  & 4595703 & 00:45 & 4596929 & 00:59 \\ \cmidrule{2-14}
			& $\mathcal{P}_{IP}^9$ &  &  &  &  & 3504258 & 00:01 &  &  & 4595703 & 00:01 & 4596929 & 00:01 \\ \midrule
			\multirow{2}{*}{10} & $\mathcal{P}_{LP}^{10}$ &  &  &  &  & 3499117 & 01:10 &  &  &  &  &  &  \\ \cmidrule{2-14}
			& $\mathcal{P}_{IP}^{10}$ &  &  &  &  & 3509608 & 00:01 &  &  &  &  &  &  \\ \midrule
			\multirow{2}{*}{11} & $\mathcal{P}_{LP}^{11}$ &  &  &  &  & 3499609 & 00:45 &  &  &  &  &  &  \\ \cmidrule{2-14}
			& $\mathcal{P}_{IP}^{11}$ &  &  &  &  & 3499609 & 00:01 &  &  &  &  &  &  \\ \midrule
			\multicolumn{2}{c}{\textbf{Final solution}} & \textbf{3502118} & \textbf{11:05} & \textbf{3501809} & \textbf{12:24} & \textbf{3499609} & \textbf{19:22} & \textbf{4594146} & \textbf{22:27} & \textbf{4595703} & \textbf{27:55} & \textbf{4596929} & \textbf{30:35} \\ \bottomrule
	\end{tabular}}
	\label{tab:IPPTerm}
	\\ \footnotesize{$^\ast$All values in the ``Cost'' columns are in USD, and all the corresponding real values are rounded-off to the next integer values. All values in the ``Time'' columns are in HH:MM, and all the corresponding seconds' values are rounded-off to the next minute values.}
\end{table}

\section{Conclusion and Future Research} \label{sec:conc}
For an airline, crew operating cost is the second largest expense, after the fuel cost, making the crew pairing optimization critical for business viability. Over the last three decades, CPOP has received an unprecedented attention from the OR community, as a result of which numerous CPOP solution approaches have been proposed. Yet, the emergent flight networks with conjunct scale and complexity largely remain unaddressed in the available literature. Such a scenario is all the more alarming, considering that the air traffic is expected to scale up to double over the next 20 years, wherein, most airlines may need to cater to multiple crew bases and multiple hub-and-spoke subnetworks. This research has proposed an Airline Crew Pairing Optimization Framework ($AirCROP$) based on domain-knowledge driven CG strategies for efficiently tackling real-world, large-scale and complex flight networks. This paper has presented not just \textit{the design of the $AirCROP$'s constitutive modules}, but has also shared insights on \textit{how these modules interact} and \textit{how sensitive the $AirCROP's$ performance is to the sources of variability, choice of different methods and parameter settings}. 
\par Given a CPOP, $AirCROP$ first preprocesses the entire duty overnight-connection network via its Legal Crew Pairing Generator\footnote{This module is utilized again to facilitate legal crew pairings when required in real-time in other modules of $AirCROP$} Subsequently, $AirCROP$ is initialized using an IFS generated by the proposed method (IPDCH). Next, the $AirCROP$'s Optimization Engine attempts to find a good-quality CPOP solution via intermittent interactions of its submodules, namely, \textit{CG-driven LPP-solutioning} and \textit{IPP-solutioning}. The efficacy of $AirCROP$ has been demonstrated on real-world airline flight network characterized by an unprecedented (in reference to available literature) conjunct scale-and-complexity, marked by over 4200 flights, 15 crew bases, multiple hub-and-spoke subnetworks, and billion-plus pairings. The distinctive contribution of this paper is also embedded in its empirical investigation of critically important questions relating to variability and sensitivity, which the literature is otherwise silent on. In that:
\begin{itemize}
    \item first, the sensitivity analysis of $AirCROP$ is performed in the presence and absence of sources of variability. It is empirically highlighted that $AirCROP$ is capable of offering comparable cost solutions, both in the presence or absence of the sources of variability. This endorses the robustness of its constitutive modules.
    \item second, the sensitivity of $AirCROP$ with respect to the cost quality of the initial solution and the associated run-time is investigated vis-$\grave{a}$-vis three different initialization methods. Again, the robustness of $AirCROP$ is endorsed, considering that it is found to be capable of offering similar cost solutions, despite the significantly varying cost and run-time of the initial solutions. 
    %
    %
    \item last, the sensitivity of $AirCROP$ to the termination parameter settings associated with the Optimization Engine’s submodules, is investigated. The fact that with the variation in termination settings of both LPP-solutioning and IPP-solutioning (independent of each other)- the $AirCROP$'s performance strongly aligns with the logically expected trends, is a testimony to the robustness of its constitutive modules.
\end{itemize} 
\par Notably, $AirCROP$ has been implemented using Python scripting language, aligned with the industrial sponsor’s preferences. However, a significant reduction in run-time could be achieved by the use of compiled programming languages such as C++, Java, etc. Moreover, employing the domain-knowledge driven CG strategies during the IPP-solutioning phase too, may augment the overall cost- and time-efficiency of the $AirCROP$. Furthermore, the emerging trend of utilizing the \textit{Machine Learning} capabilities for assisting combinatorial optimization tasks, may also hold promise for the airline crew pairing optimization, towards which an exploratory attempt has been made by the authors \citep{aggarwal2020learning}. Despite the scope for improvement, the authors hope that with the emergent trend of evolving scale and complexity of airline flight networks, this paper shall serve as an important milestone for the affiliated research and applications.

\section*{Acknowledgment}
This research work is a part of an Indo-Dutch joint research project, supported by the Ministry of Electronics and Information Technology (MEITY), India [grant number 13(4)/2015-CC\&BT]; Netherlands Organization for Scientific Research (NWO), the Netherlands; and General Electric (GE) Aviation, India. The authors thank GE Aviation, particularly, Saaju Paulose (Senior Manager), Arioli Arumugam (Senior Director- Data \& Analytics), and Alla Rajesh (Senior Staff Data \& Analytics Scientist) for providing real-world test cases, and sharing their domain knowledge which has helped the authors significantly in successfully completing this research work.

\bibliographystyle{apacite}
\bibliography{cas-refs}

\end{document}